\documentclass[aps,prb,twocolumn,superscriptaddress]{revtex4-1}
\usepackage{graphicx}
\usepackage{picture}
\usepackage{xcolor}
\usepackage{color}
\usepackage{tikz}
\usepackage{amsmath}

\begin{document}

\title{Beyond Quantum Cluster Theories: Multiscale Approaches for Strongly Correlated Systems
}

\author{Herbert F. Fotso}
\affiliation{Department of Physics, University at Albany (SUNY), Albany, New York 12222, USA}
\author{Ka-Ming Tam }
\affiliation{Department of Physics and Astronomy, Louisiana State University, Baton Rouge, LA 70803, USA}
\affiliation{Center for Computation and Technology, Louisiana State University, Baton Rouge, LA 70803, USA}
\author{Juana Moreno}
\affiliation{Department of Physics and Astronomy, Louisiana State University, Baton Rouge, LA 70803, USA}
\affiliation{Center for Computation and Technology, Louisiana State University, Baton Rouge, LA 70803, USA}

\begin{abstract}

The degrees of freedom that confer to strongly correlated systems their many intriguing properties also render them fairly intractable through typical perturbative treatments. For this reason, the mechanisms responsible for their technologically promising properties remain mostly elusive. Computational approaches have played a major role in efforts to fill this void. In particular, dynamical mean field theory (DMFT) and its cluster extension, the dynamical cluster approximation (DCA) have allowed significant progress. However, despite all the insightful results of these embedding schemes, computational constraints, such as the minus sign problem in Quantum Monte Carlo (QMC), and the exponential growth of the Hilbert space in exact diagonalization (ED) methods, still limit the length scale within which correlations can be treated exactly in the formalism. A recent advance aiming to overcome these difficulties is the development of multiscale many body approaches whereby this challenge is addressed by introducing an intermediate length scale between the short length scale where correlations are treated exactly using a cluster solver such QMC or ED, and the long length scale where correlations are treated in a mean field manner. At this intermediate length scale correlations can be treated perturbatively. This is the essence of multiscale many-body methods. We will review various implementations of these multiscale many-body approaches, the results they have produced, and the outstanding challenges that should be addressed for further advances. 

\end{abstract}

\maketitle

\section{Introduction}

Strongly correlated systems include some of the most technologically promising materials of our time. To harness their significant promise, understanding the fundamental mechanisms responsible for their intriguing properties is essential. \cite{Wolf_etal_2001,Morosan_etal_2012,Dagotto_Tokura_2008,Tokura_Nagaosa_2000,Dagotto_2005} This understanding remains a challenge for the condensed matter community despite several decades of intense effort. For instance, although the discovery of high temperature superconductors dates back to 1987,~\cite{Bednorz_Muller_1986} the underlying superconducting mechanism remains the subject of intense research activity.  Following their discovery, the Hubbard model was postulated to contain the ingredients necessary  to explain the properties of high temperature superconductors and their low-energy excitations. \cite{f_zhang_88} But despite its simplicity, an exact solution of the Hubbard model beyond one dimension remains elusive. \cite{Lieb_Wu_2003,Lieb_Wu_1968} Therefore, numerical methods have played a crucial role. These methods are however constrained by the minus sign problem for Quantum Monte Carlo (QMC), or by the exponential scaling of the Hilbert space for exact diagonalization, to relatively small system sizes. Embedding schemes have emerged as an important avenue to treat the problem in the thermodynamic limit. These schemes map the lattice problem onto an impurity, for the case of dynamical mean field theory, or onto a cluster for the case of its cluster extensions, dynamical cluster approximation or cellular dynamical mean field theory, embedded into a mean field. \cite{th_maier_05a,a_georges_96a} Embedding approaches allow the exact treatment of short length scales at the level of the cluster or the impurity, and the treatment of longer length scales at the mean field level. Multiscale many body  methods follow this logic to its natural next step by incorporating between the previous two length scales, an intermediate one at which correlations are treated with diagrammatic perturbation theory. \cite{c_slezak_06b}


In general, the difficulty in understanding correlated systems lies in the fact that there are no simple theories to explain both the weak interaction limit of the metallic state and the strong interaction limit of the Mott insulating phase.\cite{n_mott_68,n_mott_82} The most successful theory of interacting fermions is the Fermi liquid theory. \cite{Landau_1956,landau_1957,landau_1959} The basic underlying assumption is that the interaction can be turned on adiabatically from the non-interacting free fermions limit. The consequence is that the quantum numbers of the non-interacting fermions remain unchanged. Electrons can be treated as quasi-particles in a rather stable state with a lifetime that becomes very long for those states near the Fermi level. 

The Fermi liquid theory is a very efficient description of interacting fermions in a metallic phase \cite{Landau_1956,landau_1957,landau_1959}. It is  applicable to almost all metallic phases, except for special circumstances such as the notable exception of one-dimensional systems. The theory owes its simplicity to being an effective renormalized single particle theory. Once the system is beyond the simple single particle description, there is no universal prescription to handle the competition or cooperation between the different degrees of freedom and the interplay between the kinetic and the potential energies. Precisely for this reason, numerical methods are often inevitable for practical calculations. 

Widely used mean field methods factorize the interaction terms in the Hamiltonian to reduce the problem to an effective single particle theory in a static potential. The mean field, Hartree-Fock, approximation often provides reasonable results \cite{Hartree_1928,Fock_1930a,Fock_1930b,Salter_1930}, but its  shortcomings are also obvious, in particular for intermediate interaction strengths where quantum fluctuations are large. The Hartree-Fock approximation quenches the quantum or temporal fluctuations completely. This may be a reasonable assumption if the interactions are overpowered by the kinetic energy terms. However, for many physical realizations of strongly correlated systems, perhaps the most well known one being the cuprate superconductors, the interaction is of the same order of magnitude as  the bandwidth. Naively factorizing the interaction term to suppress all the quantum fluctuations is questionable at best. Indeed, there is currently no simple mean field theory that can explain most features of the cuprate superconductors. Understanding the metallic phases beyond Fermi liquid theory is key for understanding broken symmetry phases, such as d-wave superconducting pairing in the cuprates. While one can construct a phase with no explicit broken symmetry and use the mean field method to understand the effective theory, this always involves fractionalized particles and strong constraints such as those of gauge theories. \cite{Lee_etal_2006}

Beyond mean field theory, there exists a plethora of techniques based on weak coupling expansion. They are typically based on low order perturbative methods, such as second order perturbation theory, or on selecting a certain class of diagrams and summing them up to infinite order. A typical example is the random phase approximation (RPA), which selects the class of ladder diagrams and sums them up to infinite order. \cite{Bohm_Pines_1951,Bohm_Pines_1952,Bohm_Pines_1953,Gell-Mann_Brueckner_1957} A more sophisticated approach is to sum a large class of diagrams in an iterative way. 
For example, parquet diagrams are generated when second order diagrams are inserted iteratively into the interaction vertex. This generates a class of diagrams that can only be separated into two disconnected pieces by cutting at least two fermion lines. \cite{Landau_1954a,Landau_1954b,Landau_1954c}
The advantage of the parquet approach compared to second order perturbation theory is in the ability to sum up a large variety of diagrams including those at infinite order. This, in principle, allows the instability towards a broken symmetry to be captured \cite{n_bickers_98,n_bickers_91,n_bickers_93}. Its main advantage over random phase approximation is in its unbiased sum of diagrams in different scattering channels to enable the study of the competition among different broken symmetries.

Instead of using a diagrammatic expansion approach, the Dynamical Mean Field Theory (DMFT) maps the strongly correlated lattice onto an impurity site embedded in a self-consistently determined effective medium. 
The interest in the high spatial dimension limit of strongly correlated models in the late 80's and early 90's led to the understanding that in this limit, 
strongly correlated models with local interactions can be greatly simplified. This is due to the fact that an expansion in  terms of the hopping amplitude in infinite dimension leads to the vanishing of all diagrams except the local ones, and, for a translationally invariant system, the model loses all spatial dependence. This simplification led to the dynamical mean field theory.

DMFT remains  the subject of active research efforts, particularly because there is no universal quantum impurity solver. Various methods have been proposed over the past few decades. These include semi-analytical methods based on perturbation theory or modified mean field theories. The more well known methods include the iterative perturbation theory and the local moment approximation. Numerical approaches include various kinds of Quantum Monte Carlo and exact diagonalization methods. Recently, density matrix renormalization group and matrix product state methods have also been explored. 
Quantum computing algorithms for solving the quantum impurity problem have been proposed recently. \cite{Bauer_etal_2016,Rungger_etal_2020,Keen_etal_2019,McClean_etal_2016}  After all, solving even a single impurity is a non-trivial problem, as the mean field hybridization function is not given by a simple form that can allow an analytical solution. 

It is worth noting that the DMFT can be viewed as a formal generalization of the coherent potential approximation (CPA) proposed by Soven in the 60's. \cite{soven_cpa,shiba71} The CPA has since been extensively used for studies of random disorder models with negligible interactions, in particular for random alloys. 
Various extensions of CPA have been proposed over the years. The earliest one that goes beyond the single site approximation is the molecular CPA. \cite{Lempert_Hass_Ehrenreich_1984,Lempert_Hass_Ehrenreich_1987}  It embeds the cluster into an effective medium that possesses the same structure as the cluster itself. It thus generates an effective medium different from that of the CPA. The obvious deficiency of the method is the breaking of translational invariance.

A different scheme of embedding cluster methods is the Dynamical Cluster Approximation (DCA)~\cite{m_hettler_00a,m_hettler_98a}. By construction, the method naturally preserves the translational invariance of the original model by directly working with both the cluster and the effective medium in momentum space. The method has been extensively employed on the Hubbard model \cite{Gutzwiller_1963,Kanamori_1963,Hubbard_1963}, Anderson model \cite{Anderson_1958}, periodic  Anderson model \cite{Anderson_1961}, and Falicov-Kimball model \cite{Falicov_Kimball_1969}. The cluster extension is not just a quantitative improvement on the DMFT. It is necessary to produce important features that are absent in the DMFT results. Perhaps the most important one is the DCA's ability to capture nonlocal correlations such as that of d-wave pairing, which is obviously absent for approximations that do not consider spatial dependence explicitly\cite{Tsuei_1994}.  The method has also been considered in the context of multiple scattering theory where it is re-branded as non-linear coherent potential approximation\cite{Rowlands_2003}.

The difficulty of solving a cluster impurity (or embedded cluster) problem scales exponentially.  Roughly speaking, quantum Monte Carlo based methods scale exponentially with the number of impurity sites (cluster size), with inverse temperature, or with the interaction strength \cite{j_hirsch_86,e_gull_08,e_gull_11,a_rubtsov_05}. The exception is strong-coupling expansion based Monte Carlo methods, but this is usually limited to a rather small number of impurity sites\cite{p_werner_06a,p_werner_06b}. Another class of impurity/cluster solvers is based on diagonalization of the effective finite size Hamiltonian. For these Hamiltonian-based solvers, such as exact diagonalization \cite{Liebsch08,Senechal_2008,liebsch2011temperature,Gunnarsson_Schonhammer_1983,Hong_Kee_1995,e_koch_08}, the Hilbert space grows exponentially with the cluster size and thus both computing memory and time requirements grow at the same rate. This is also true to a large extent for another Hamiltonian-based approach, the numerical renormalization group method \cite{NRG_rmp,h_krishnamurthy_80a,h_krishnamurthy_80b,k_wilson_75}. In general, for practical calculations, the maximum number of impurity sites is rather modest ($\sim10$ sites). 

Over the last couple of decades various novel methods have been proposed. These include the density matrix renormalization group \cite{Fernandez_Garcia_Hallberg_2014,Zhe_etal_2017}, the related matrix product wavefunction \cite{Wolf_etal_2014a} and, even more recently, different forms of machine learning approaches \cite{Arsenault_2014,Walker_2020}. These more recent methods may have potential for certain applications. For instance, they may be more efficient for calculating real time Green functions in nonequilibrium problems \cite{Wolf_etal_2014b}. Approaches based on machine learning could also be more efficient in solving a large set of impurity problems, and this may be useful for applications on random systems that require averaging over random disorder realizations \cite{Terletska_etal_2018}. However, none of these novel impurity solvers are suitable for the calculation of the vertex function which is essential for most methods that are built on an expansion on top of the DMFT solution. Additionally, Monte Carlo sampling of the partition function provides more flexibility for controlling the error as the impurity cluster size is increased. Also, although it has been proven that the single impurity problem does not exhibit a minus sign problem, the absence of minus sign in the Monte Carlo sampling can not be assumed for a  generic impurity problem  \cite{Yoo_etal_2005}.



Following the logic of embedding a small system into a mean-field host, one can anticipate better accuracy in the result if an intermediate length scale is inserted between the previous two. Since short length scales are appropriately treated with exact solvers and the long length scale by a mean field, this intermediate length scale can be treated reliably with diagrammatic methods. This is the essence of multiscale many body methods for strongly correlated systems as formulated in the early 2000's \cite{c_slezak_06b} and the subject of continued efforts since then \cite{Rohringer_etal_2018}.

This review focuses exclusively on the methods for studying systems in equilibrium. It is noteworthy to point that effort has been devoted to the generalization of the multiscale many body (MSMB) approaches to non-equilibrium problems \cite{Jung_etal_2012, Zhou_Guo_2019, Chen_etal_2019}. We refer the reader to the original article for  details \cite{Jung_etal_2012}, as we do not discuss these nonequilibrium approaches in the present paper. 

The rest of the review is structured as follows. In section \ref{sec:DMFT}, we summarize the DMFT method and its connection to the Anderson impurity problem by approaching it from its "cavity method" formulation. In section \ref{sec:Clusters}, we discuss two cluster extensions of DMFT, dynamical cluster approximation (DCA) and cellular dynamical mean field theory (CDMFT). We proceed  in section \ref{sec:EDMFT} with a discussion of the extended dynamical mean field theory (EDMFT) that extends DMFT to the treatment of nonlocal interaction. Section \ref{sec:1-over-d} is focused on the $1/d$ expansion, a systematic expansion of DMFT with respect to the hopping amplitude. In section \ref{sec:MSMB}, we describe the original formulation of the multiscale many body method. The parquet formalism, which encompasses various commonly used diagrammatic approximations, and which is essential for the diagrammatic treatment of intermediate length scales, is described in section \ref{sec:Parquet}. Following this, we briefly describe different implementations  to incorporate nonlocal corrections into the DMFT/DCA  starting with the dynamical vertex approximation in section \ref{sec:DGammaA}, and then the dual fermion method in section \ref{sec:DualFermions}.  In section  \ref{sec:DualBosons}, we present the dual boson extension for corrections to EDMFT. In section \ref{sec:GW}, we discuss efforts to incorporate nonlocal correction into the GW approximation (an approach that obtains the self-energy from the single particle Green function (G) and the screened Coulomb interaction (W)) in the form of the ``triply irreducible local expansion" (TRILEX). In section \ref{sec:FRG}, we discuss functional renormalization group and its usage for nonlocal corrections to DMFT. In section \ref{sec:Numerics}, an important computational challenge, the numerical representation of the vertex functions in memory, is discussed. In section \ref{sec:Constraints}, we discuss important physical constraints on the methods. In section \ref{sec:Applications}, we summarize results obtained on different models with various implementations of the multiscale many-body approach  before ending with our conclusions.

\section{Dynamical Mean Field Theory}
\label{sec:DMFT}

\begin{figure}[htbp]
\includegraphics*[width=5.0cm, height=5.20cm]{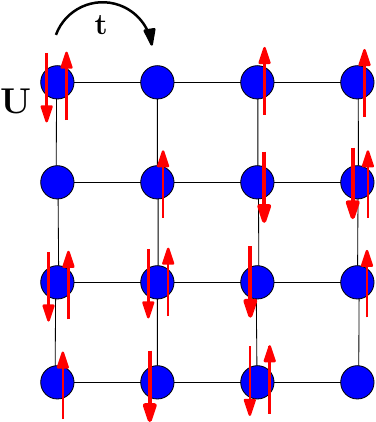} 
\caption{Schematic illustration of the Hubbard model on a square lattice. Electrons are allowed to hop between nearest neighboring sites with a hopping amplitude $t$, and the interaction of strength $U$ is local.
} 
\label{fig:HubbardLattice}
\end{figure}

Following the discovery of high temperature superconductors, it was argued that the Hubbard model (\ref{eq:Hubbard}) captures their low energy properties. \cite{f_zhang_88} This deceptively simple model describes itinerant electrons that can hop between nearest-neighbor sites $\langle i,j \rangle $ on a lattice with a hopping integral $t_{ij}$, and are subject to a Coulomb interaction $U$ when a site is doubly occupied. The model is schematically depicted in Fig. \ref{fig:HubbardLattice} and defined as:
\begin{eqnarray}
\hat{H} = - \sum_{<i,j>,\sigma} t_{ij} c^\dagger_{i,\sigma} c_{j,\sigma} + \sum_i U \hat{n}_{i,\uparrow} \hat{n}_{i,\downarrow},
\label{eq:Hubbard}
\end{eqnarray}
where $c_{i, \sigma}$ $(c_{i,\sigma}^{\dagger})$ is the destruction (creation) operator that destroys (creates) an electron with spin $\sigma$ at site $i$. $\hat{n}_{i,\sigma} = c^\dagger_{i,\sigma} c_{i,\sigma}$ is the number of particles of spin $\sigma$ at site $i$.

The DMFT method in itself is a generalization of the usual mean field theory. But unlike the usual mean field theory, say for the Ising model for example, the mean field here is not an order parameter. Instead it is a function of time or frequency. Thus, the approach captures all temporal fluctuations. Focusing on the translationally invariant paramagnetic phase, there is no explicit order parameter.  One may also construct an explicit order parameter to represent the broken symmetry, but this is not done routinely and is not the focus of this discussion. 


Within the DMFT solution of a translationally invariant system, the spatial fluctuations are completely suppressed as in a traditional mean field theory.
As we will see, a major way to improve the approximation is to systematically incorporate the corrections due to the spatial dependence of the model into the DMFT solution. Also, since 
the self-consistency condition is valid only at the single particle level, there is no guarantee that high order Green functions, such as the different susceptibilities, are matched between the lattice and impurity models. 

When the  DMFT was developed in the early 90's \cite{w_metzner_89a,w_metzner_89b,e_mullerhartmann_89a,e_mullerhartmann_89b,a_georges_92a,m_jarrell_92a,a_georges_96a,v_janis_91,Metzner_Vollhardt_1990}, solving even a single impurity problem was rather challenging. With the formulation of new numerical algorithms and advances in computing power, a single impurity problem can, in general, be numerically solved quite efficiently. While there is still not a completely satisfactory method that is accurate for a wide range of parameters, particularly for solving interacting problems with random disorder which requires a large ensemble of different impurity realizations for disorder averaging
\cite{Terletska_etal_2018,v_dobrosavljevic_03,c_ekuma_15c}, the single impurity problem can mostly be handled at the present time. 

As a side note, it is worth mentioning that earlier work on
DMFT can be traced back to the study of the transverse-field Sherrington-Kirkpatrick model.\cite{Sherrington_Kirkpatrick_1975} In that model the spatial fluctuations are completely suppressed as the model consists of spin couplings in the fully connected network with the transverse external magnetic field, thus it can be mapped to a single site problem. \cite{Bray_Moore_1980} The idea of similarly handling fermion problems only appeared in the late 80's in studies of fermionic systems in the infinite dimensional limit by Vollhardt and Metzer. \cite{w_metzner_89a,w_metzner_89b} A series of papers by Müller-Hartmann were also influential in the development of DMFT and later generalizations to the dynamical cluster approximation. \cite{e_mullerhartmann_89a,e_mullerhartmann_89b} In 1991, the physics of the Hubbard model via DMFT was discussed by Georges and Kotliar. \cite{a_georges_92a} The first numerically ``exact'' DMFT solution of the Hubbard model was presented by Jarrell. \cite{m_jarrell_92a}

The DMFT formalism can be explained quite transparently from a path integral formulation. Consider the action of the Hubbard model on a lattice, 
\begin{eqnarray}
&&S = - \sum_{\mathbf{r}_{i},\mathbf{r}_{j},\sigma} \int_{0}^{\beta}   \int_{0}^{\beta}  d\tau_{i} d\tau_{j} \psi_{\sigma}^{*}(\mathbf{r}_i,\tau_{i}){G}_{0}^{-1}(\mathbf{r}_i,\tau_{i},\mathbf{r}_j,\tau_{j}) \psi_{\sigma}(\mathbf{r}_j,\tau_{j}) \nonumber \\   
&+& U \sum_{\mathbf{r}_{i}}\int^{\beta}_{0}d\tau_{i} \psi_{\uparrow}^{*}(\mathbf{r}_i,\tau_{i})\psi_{\uparrow}(\mathbf{r}_{i}, \tau_{i})\psi^{*}_{\downarrow}(\mathbf{r}_{i},\tau_{i})\psi_{\downarrow}(\mathbf{r}_{i},\tau_{i}),
\label{eq:action1}
\end{eqnarray}
where $G_{0}$ is the bare Green function, $\psi_{\sigma_{i}}^{*}(\mathbf{r}_i,\tau_{i})$ and $\psi_{\sigma_{i}}(\mathbf{r}_i,\tau_{i})$ are the Grassmann fields for spin $\sigma_{i}$ at location $\mathbf{r}_i$ and imaginary time $\tau_{i}$.

The first part of the action contains the kinetic energy as characterized by the bare Green function $G_0$. It is simply obtained from the bare dispersion of the considered model. The second term includes the interaction characterized by the parameter $U$, which we assume to be  local. Any interaction beyond the local Hubbard term will involve further approximations in the context of the dynamical mean field theory.

The exact Green function of the above action can be completely characterized by the self-energy $\Sigma$. If we write the self-energy in the frequency-momentum space, the relation between the bare Green function and the exact Green function $G$ is given by the Dyson equation,

\begin{eqnarray}
    G(\mathbf{k},\omega) = \frac{1}{G_{0}^{-1}(\mathbf{k},\omega)-\Sigma(\mathbf{k},\omega)}.
\end{eqnarray}

In the simplified case of a translationally invariant system, the idea of the dynamical mean field theory is to relate 
the full lattice problem with spatial dependence to a single site problem. To this end, DMFT reframes (\ref{eq:action1}) into an effective action for a single site with a bare Green's function $\mathcal{G}$: 
\begin{eqnarray}
\label{eq:actionEff1}
&&S_{eff} =  - \int d\omega \sum_{\sigma} \psi_{\sigma}^{*}(\omega)\mathcal{G}^{-1}(\omega) \psi_{\sigma}(\omega) \\
&+& U \int_{\omega_1+\omega_3=\omega_2+\omega_4} d\omega_1 d\omega_2 d\omega_3 d\omega_4 \psi_{\uparrow}^{*}(\omega_1)\psi_{\uparrow}( \omega_2)\psi^{*}_{\downarrow}(\omega_3)\psi_{\downarrow}(\omega_4).
\nonumber 
\end{eqnarray}
Consider the Anderson impurity model characterizing an impurity coupled to a band of conducting electrons and given by the Hamiltonian:
\begin{eqnarray}
 \label{eq:AIMhamiltonian}
 H_{AIM} & = & U n_{\uparrow} n_{\downarrow} - \mu \sum_{\sigma} c_{\sigma}^{\dagger} c_{\sigma} \nonumber \\
 & + & \sum_{j, \sigma} V_j \left( f_{j \sigma}^{\dagger}c_{\sigma} +  c_{\sigma}^{\dagger} f_{j \sigma} \right) + \sum_{j \sigma} \epsilon_j f_{j \sigma}^{\dagger} f_{j \sigma}.
\end{eqnarray}
Where $c_{\sigma}^{\dagger}$, $c_{\sigma}$ are creation and destruction operators for the impurity electrons while $f_{j \sigma}^{\dagger}$, $f_{j \sigma}$ are those of the conduction electrons.
Its action is:
\begin{eqnarray}
 \label{eq:AIMaction}
 & &S_{AIM} =  - \int d\omega \sum_{\sigma} \psi_{\sigma}^{*}(\omega)\mathcal{G}^{-1}_{AIM}(\omega) \psi_{\sigma}(\omega) \\
&+& U \int_{\omega_1+\omega_3=\omega_2+\omega_4} d\omega_1 d\omega_2 d\omega_3 d\omega_4 \psi_{\uparrow}^{*}(\omega_1)\psi_{\uparrow}( \omega_2)\psi^{*}_{\downarrow}(\omega_3)\psi_{\downarrow}(\omega_4),
\nonumber 
\end{eqnarray}
where $\mathcal{G}_{AIM}$ is the single impurity Anderson model non-interacting Green's function defined by:
\begin{equation}
 \mathcal{G}^{-1}_{AIM}(i\omega_n) = i\omega_n + \mu - \Delta(i\omega_n)
 \label{eq:GscriptAIM}
\end{equation}
with
\begin{equation}
 \Delta(i\omega_n)= \int_{-\infty}^{+\infty} d\omega \frac{1}{i\omega_n - \omega} \sum_{j \sigma} V_j^2 \delta(\omega - \epsilon_j).
\end{equation}
The AIM action (\ref{eq:AIMaction}) is equivalent to (\ref{eq:actionEff1}) with $\mathcal{G}$ playing the role of the non-interacting AIM Green's function. The construction can be justified via the concept of the "cavity method" in the infinite dimension limit whereby all degrees of freedom are integrated out except for the site labelled by the index $0$. In this limit of $d \to \infty$ for a hypercubic lattice, the hopping amplitude is rescaled as  $t_{ij} = t/\sqrt{2d}$ so that the kinetic energy and the interaction energy remain of the same order. The effective action, $S_{eff}$, in this process is defined by:
\begin{eqnarray}
\frac{1}{Z_{eff}}exp(-S_{eff}[\psi_{0,\sigma}^{*}\psi_{0,\sigma}])=  \nonumber \\
\frac{1}{Z}\int \prod_{i \neq 0,\sigma}exp(-S[\{\psi_{i,\sigma}^{*},\psi_{i,\sigma}\}]).
\label{eq:actionEff2}
\end{eqnarray}
Where $Z$ and $Z_{eff}$ are the partition functions associated with $S$ and $S_{eff}$ respectively. 
The effective action can be written as:
\begin{eqnarray}
&&S_{eff}[\psi_{0,\sigma}^{*}\psi_{0,\sigma}] =  S_0 \nonumber \\  &+& \sum_{n=1}^{N} \sum_{i_{1},j_{1},\cdot\cdot\cdot i_{n},j_{n}} 
\eta^*_{0,i_{1}} \eta^*_{0,i_{2}}  \cdot\cdot\cdot \eta^*_{0,i_{n}}
\eta_{j_{1},0}\eta_{j_{2},0} \cdot\cdot\cdot \eta_{j_{n},0}  \nonumber \\ \nonumber  &&\prod_{i,j=1,\cdot\cdot\cdot,n} \int d\tau_i d\tau_j 
 G^{(0)}(i_1,\tau_{i_1}, i_2, \tau_{i_2},\cdot\cdot\cdot,i_n,\tau_{i_n}; \\  &&j_1,\tau_{j_1}, j_2, \tau_{j_2},\cdot\cdot\cdot,j_n,\tau_{j_n}). 
 \label{eq:actionEff3}
\end{eqnarray}
Where $S_0$ is the local action at site "0":
\begin{eqnarray}
S_{0} &=& - \int_{0}^{\beta}   \int_{0}^{\beta}  d\tau d\tau^{'} \sum_{\sigma} \psi_{\sigma}^{*}(\tau){G}_{0}^{-1}(\tau,\tau^{'}) \psi_{\sigma^{}}^{}(\tau^{'}) \nonumber  \\  
&+& U \int^{\beta}_{0} d\tau \psi_{\uparrow}^{*}(\tau)\psi_{\uparrow}( \tau)\psi^{*}_{\downarrow}(\tau)\psi_{\downarrow}(\tau). 
\label{eq:dmft_s0}
\end{eqnarray}
$G^{(0)}$ is the Green's function connecting the cavity to the impurity. $\eta_{i,0}= t_{i,0} \psi_{0,\sigma}$, with $t_{i,0}$ the hopping from site $i$ to $0$.

Only terms of order $n=2$ survive the expansion $(\ref{eq:actionEff3})$ in the $d\to \infty$ limit. Leading to:
\begin{eqnarray}
&&S_{eff}[\psi_{0,\sigma}^{*}\psi_{0,\sigma}] = S_0 \nonumber \\ &&  + \sum_{i_1, j_1=1}^{N}  t_{0,i_{1}} t_{j_{1},0}\int d\tau d\tau^{'} \psi_{0,\sigma}^{*}\psi_{0,\sigma} G(i_1,\tau_{i_1}, j_1,\tau_{j_1}).
\end{eqnarray}

Rewriting this in frequency space, gives the AIM action (\ref{eq:AIMaction}) with $\mathcal{G}_{AIM}$ replaced by $\mathcal{G}$ such that:
\begin{eqnarray}
\mathcal{G}^{-1}(i\omega_n) = i\omega_{n}  + \mu - \sum_{i,j} t_{0,i}t_{j,0} G(i\omega_{n}).
\end{eqnarray}
This relation connects the impurity Green function to the lattice Green function. For the Bethe lattice, \cite{Bethe_1935} $\mathcal{G}^{-1}(i\omega_n) = i\omega_{n}  + \mu  - t^{2} G(i\omega_{n})$. 
For a general lattice, the connection between the lattice Hubbard model and the single impurity model is established by setting the self-energy, 
\begin{eqnarray}
    \Sigma_{lattice}(\mathbf{k},\omega) = \Sigma_{impurity}(\omega).
\end{eqnarray}

Since the self-energy of the original lattice Hubbard model has spatial dependence while that of the single impurity Anderson model does not, to construct the lattice Green function one has to rely on the coarse-graining process that assumes the self-energy of the lattice model to be the same in the entire Brillouin zone.

\begin{eqnarray}
    G_{lattice}(\mathbf{k},\omega) = \frac{1}{G_{lattice,0}^{-1}(\mathbf{k},\omega)-\Sigma(\mathbf{k},\omega)} \\ \nonumber
    \approx \frac{1}{G_{lattice,0}^{-1}(\mathbf{k},\omega)-\Sigma_{lattice}(\omega)}.
\end{eqnarray}

The missing link between the Hubbard model and the Anderson model is to determine the effective bare Green function of the Anderson model. For the $D$-dimension case, this is given by: 
\begin{eqnarray}
    G_{impurity}(\omega) = 
    \sum_{\mathbf{k}} \frac{1}{(2\pi)^{D}} G_{lattice}(\mathbf{k},\omega) \\ \nonumber
   = \sum_{\mathbf{k}}
   \frac{1}{(2\pi)^{D}}
    \frac{1}{G_{lattice,0}^{-1}(\mathbf{k},\omega)-\Sigma_{lattice}(\omega)}.
\end{eqnarray}

\begin{figure}[htbp]
\includegraphics*[width=10.0cm, height=6.0cm]{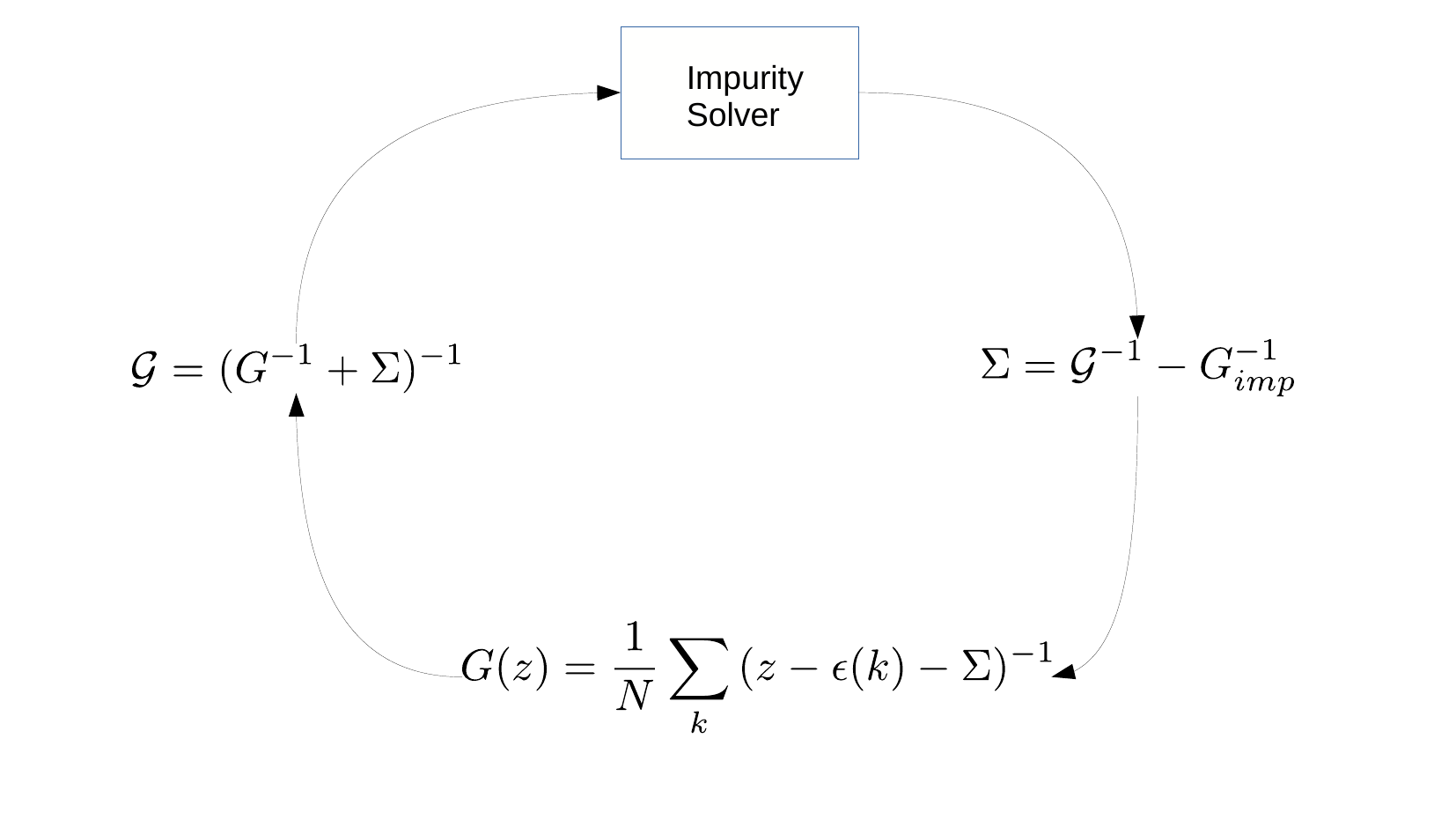} 
\caption{DMFT algorithm: {\bf Step 1.} The impurity solver provides the impurity Green function, $G_{imp}$. {\bf Step 2.} The impurity Dyson equation is used to extract the self-energy $\Sigma$. {\bf Step 3.} The self-energy is coarse-grained over the entire first Brillouin zone to obtain the lattice Green function. {\bf Step 4.} The bath Green function for the impurity problem is calculated and fed into the impurity solver. {\bf Step 5.} Repeat steps 1 to 4 until convergence is obtained for the self-energy or Green function. 
} 
\label{fig:DMFT_algorithm}
\end{figure}

The summation over the momentum can thus be replaced by an integral over the bare density of states to simplify the calculation. The bare density of states for the hypercubic lattice at the limit of infinite spatial dimensions can be exactly calculated.  \cite{e_mullerhartmann_89a,e_mullerhartmann_89b} We have gathered all the ingredients for the DMFT approximation and the algorithm can be summarized as in Fig. \ref{fig:DMFT_algorithm}.

\section{Cluster Route for extending the DMFT}
\label{sec:Clusters}

A natural and direct avenue to generalize the DMFT is to incorporate nonlocal correlations by including more than a single impurity site, i.e by formulating the theory around a cluster of multiple sites in a self-consistently determined host. This type of cluster DMFT remains an important method for the study of strongly correlated systems, as it allows, by increasing the cluster size, a systematic correction unlike perturbative expansion methods. Moreover, one can envision that a perturbative expansion on top of the cluster method would produce an even better result, since the bare effective Hamiltonian or action for the perturbative expansion, which corresponds to the DCA or CDMFT solution on the smaller system, is presumably more accurate and already includes a substantial amount of nonlocal correlations.  

For the classical spin model, the first attempt of a multiple site mean field theory was the so-called Bethe-Peierls-Weiss approximation using a cluster of $(z+ 1)$ sites, with one site at the center surrounded by $z$ sites on the shell. \cite{Weiss_1948,Bethe_1935,Peierls_1936,Kikuchi_1951} The interaction between the center spin and its $z$ nearest neighbor spins is treated explicitly while the interaction between the remaining $z$ spins with the other spins outside their own cluster is treated by a mean field. 

Another approach by Oguchi is a more direct generalisation of the mean field method. \cite{Oguchi_1955} A cluster of $N_c$ spins is considered. The interaction among these $N_c$ spins is treated explicitly, and the interaction between the spins at the edge of the cluster and spins outside of the cluster is treated by a mean field. Unlike the Bethe-Peierls-Weiss method where pairwise interactions are treated explicitly only for pairs involving the central spin, pairwise interactions among all spins of the $N_c$ spin cluster are treated explicitly in the Oguchi approach. 

A cluster of impurities in  real space is considered instead of a single one for the cellular dynamical mean field theory\cite{g_kotliar_01,Lichtenstein_Katsnelson_2000}.  There is a technical problem with using such an approach for the paramagnetic solution of the Hubbard model as the cluster naturally breaks translational invariance. A procedure for restoring the symmetry is needed for a periodic solution. \cite{Biroli_2004} 

A further approach for a cluster generalization of DMFT is based on the idea of coarse-graining that is central to the dynamical cluster approximation (DCA). \cite{k_aryanpour_02a,h_fotso_12,m_hettler_00a,m_jarrell_01c} 
A cluster of impurities is used, but after the local cluster is solved the lattice quantities are averaged over different patches of the first Brillouin zone for a coarse-grained quantity. The advantage is that the method is manifestly translationally invariant. This allows a perturbative expansion to be implemented on top of the DCA solution more naturally. As we will discuss, almost all perturbative expansion methods become simpler and less cumbersome to implement when formulated in momentum space.

One can consider that the DMFT impurity bare Green function is the coarse-grained lattice Green function. Since the self-energy which is obtained by solving the impurity model does not have spatial dependence, the coarse-graining procedure for DMFT is done over the entire first Brillouin zone with one single impurity site.  Effectively, from a diagrammatic point of view, DMFT neglects momentum conservation at the internal vertices. The dynamical cluster approximation systematically restores momentum conservation at these internal vertices. To this end, it divides the Brillouin zone into $N_c$ cells with each cell (of linear size $\Delta k$) represented by a cluster momentum \textbf{K} in the center of the cell.  The DCA then requires that momentum conservation in the internal vertices be respected for momentum transfers between cells (momentum transfers larger than $\Delta k$), but neglected for momentum transfers within a cell (less than $\Delta k$). In this way momentum conservation is fully recovered in the limit of $N_c \to \infty$, while the DMFT result is obtained for $N_c = 1$.

\begin{figure}[htbp]
\includegraphics*[width=8.0cm, height=8.0cm]{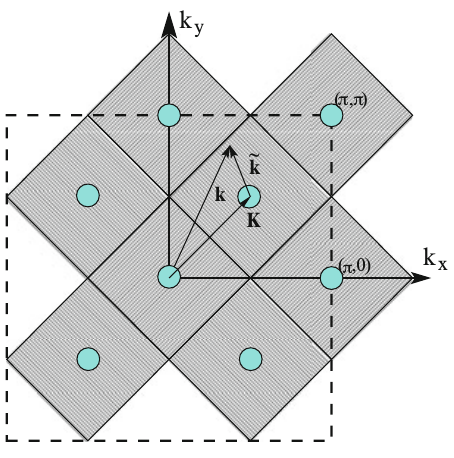} 
\caption{Illustration of the coarse-graining process in DCA for an 8-site cluster. The momentum $\mathbf{k}$ is mapped onto the nearest cluster point $\mathbf{K}$ so that $\tilde{\mathbf{k}} = \mathbf{k} - \mathbf{K}$ remains inside the cell containing $\mathbf{K}$.
} 
\label{fig:coarseGraining}
\end{figure}

The DCA coarse-graining process is illustrated by Fig.(\ref{fig:coarseGraining}) for $N_c = 8$. In DCA, the self-energy is no longer momentum-independent. Rather, we have for a lattice momentum $\mathbf{k}$, $\Sigma(\mathbf{k}) = \Sigma(\mathbf{K})$. Where $\mathbf{K}$ is the momentum at the center of the cell containing $\mathbf{k}$. The impurity, or in this case the cluster, Green's function is here related to the lattice Green's function by:
\begin{eqnarray}
    G_{impurity}(\mathbf{K},\omega) \approx 
    \sum_{\mathbf{k}} \frac{1}{(2\pi)^{D}} G_{lattice}(\mathbf{k},\omega) && \\ \nonumber
   = \sum_{\mathbf{k}}
   \frac{1}{(2\pi)^{D}}
    \frac{1}{G_{0,lattice}^{-1}(\mathbf{k},\omega)-\Sigma_{lattice}(\mathbf{K},\omega)},
\end{eqnarray}
where the summation over $\mathbf{k}$ is restricted to the patch corresponding to the impurity/cluster site with momentum $\mathbf{K}$. The impurity solver is now a cluster solver for the momentum-dependent self-energy $\Sigma(\mathbf{K})$.

\begin{figure}[htbp]
\includegraphics*[width=9.0cm, height=5.5cm]{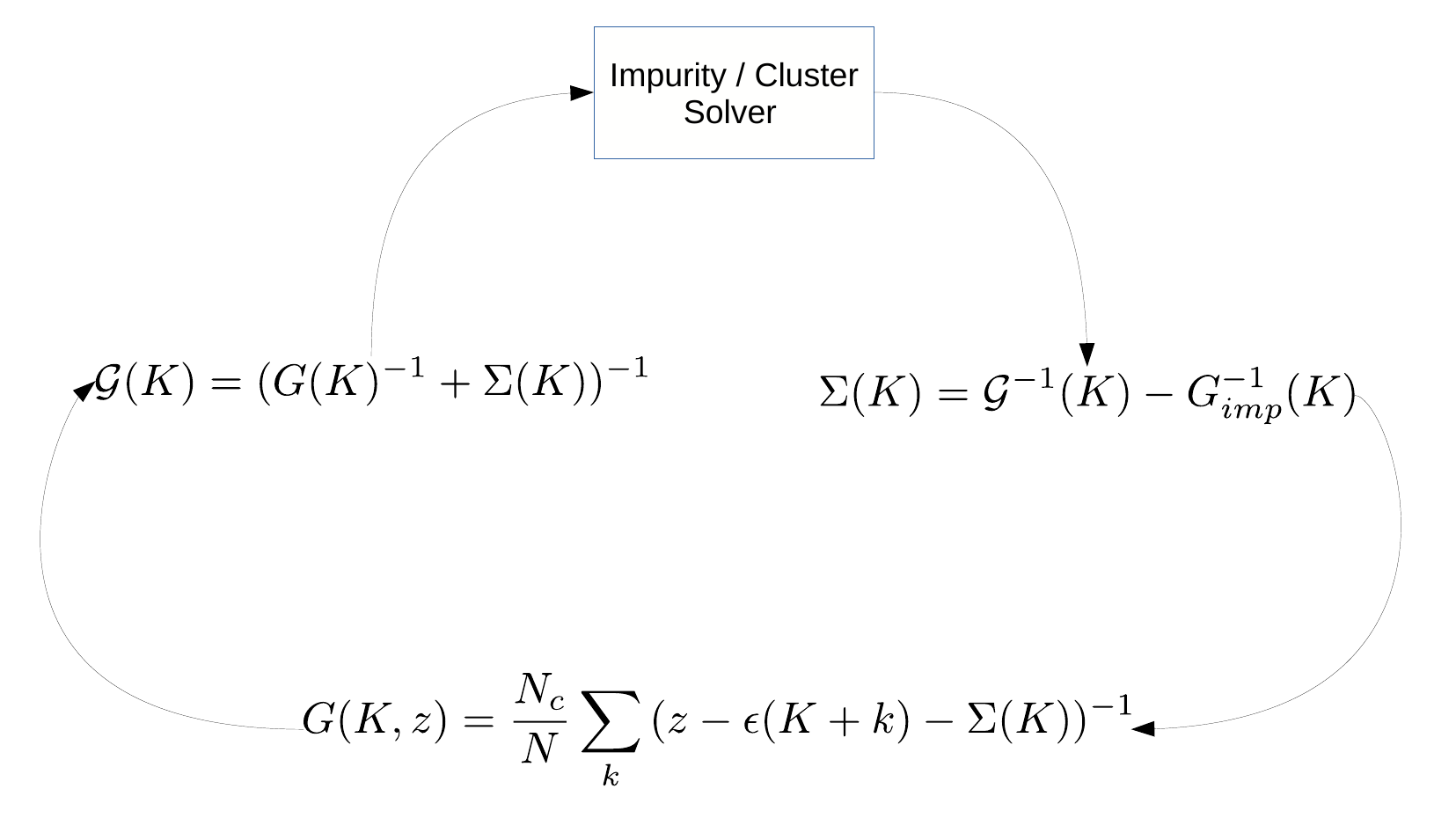} 

\caption{DCA algorithm: {\bf 1.} The impurity solver provides the impurity Green function, $G_{imp}(K)$. {\bf 2.} The Dyson equation is used to extract the self-energy $\Sigma(K)$. {\bf 3.} The self-energy is coarse-grained over the patches that fill the first Brillouin zone to obtain the lattice Green function, $G(K)$ {\bf 4.} The bath Green function, $\mathcal{G}(K)$, for the impurity problem is feed into the impurity solver. {\bf 5.} Repeat from step 1 until convergence is obtained for the self-energy or Green function. 
} 
\label{fig:DCA_algorithm}
\end{figure}

The algorithm for DCA is summarized in Fig. (\ref{fig:DCA_algorithm}).


\section{Extended Dynamical Mean Field Theory (EDMFT)}
\label{sec:EDMFT}

The earliest attempt to include the effect of nonlocal interactions was motivated by the competition between the local interaction and the RKKY interaction in heavy fermion materials. \cite{Si_Smith_1996} The basic idea was to include the density-density and/or spin-spin interaction term and scale the interaction strength with respect to the spatial dimensionality so that its fluctuations are non-zero in the high dimension limit. 
 \cite{Smith_Si_2000,Si_Smith_1996,Smith_Si_1999,Chitra_Kotliar_2000,Lleweilun_etal_2000,Sun_Kotliar_2002} This procedure leads to an impurity embedded in a self-consistent fermionic bath and, at the same time, a self-consistent bosonic bath due to the nonlocal interactions. 

In DMFT, the hopping is considered as a function of spatial dimension $d$, $t_{ij} \sim 1/\sqrt{d}^{|i-j|}$ in the infinite dimension limit. The EDMFT includes the nonlocal interaction in a similar manner: $V_{ij} \sim 1/\sqrt{d}^{|i-j|}$. \cite{Smith_Si_2000,Si_Smith_1996,Smith_Si_1999}  Here, we only consider the density-density interaction. A generic two-body interaction can include spin-spin, correlated hopping, and pair hopping. For example, if the nonlocal interaction is density-density interaction, such as in the extended Hubbard model, the effective action for the impurity problem acquires an extra term including a retarded density-density coupling mediated by the charge susceptibility. \cite{Si_Smith_1996,Chitra_Kotliar_2000}

\begin{eqnarray}
S_{0} &=&  \int d\tau d\tau^{'} \sum_{\sigma} \psi_{\sigma}^{*}(\tau){G}_{0}^{-1}(\tau,\tau^{'}) \psi_{\sigma^{}}^{}(\tau^{'}) \nonumber \\
\nonumber 
&-& U \int d\tau n_{\uparrow}(\tau)n_{\downarrow}(\tau) \\ \nonumber
&+&\int d\tau d\tau^{'}  \phi^{*}(\tau){D}_{0}^{-1}(\tau,\tau^{'}) \phi(\tau^{'}) \\
&+&\int d\tau \phi(\tau)n(\tau).
\end{eqnarray}

The effective action is equivalent to that of the single impurity model where $\mathcal{G}$ and $\mathcal{D}$ act respectively as  effective fields for the fermionic bath and for the bosonic bath in the Anderson impurity model. They can be obtained as usual via the Dyson equations. For the Bethe lattice, \cite{Chitra_Kotliar_2000}
\begin{eqnarray}
\mathcal{G}^{-1}(i\omega_n) = i\omega_{n}  + \mu - \sum_{i,j} t_{0,i}t_{j,0} G(i\omega_{n}),
\end{eqnarray}

and 
\begin{eqnarray}
\mathcal{D}^{-1}(i\omega_n) =
\sum_{i,j} V_{0,i}^{-1}V_{j,0}^{-1} D (i\omega_{n}).
\end{eqnarray}

Since the retarded coupling can be understood in term of a bosonic field coupled to the impurity charge density, EDMFT remedies the limitation of DMFT where the interaction is strictly restricted to the local on-site Hubbard interaction. We note that the decoupling of the interaction by Hubbard-Stratonovich fields is not unique: one can decouple the coupling in different ways. \cite{Ayral_etal_2012} The decoupling presented above only decouples the density-density interaction, the $V$ term in the extended Hubbard model.

\begin{figure}[htbp]
\includegraphics*[width=10.0cm, height=6.0cm]{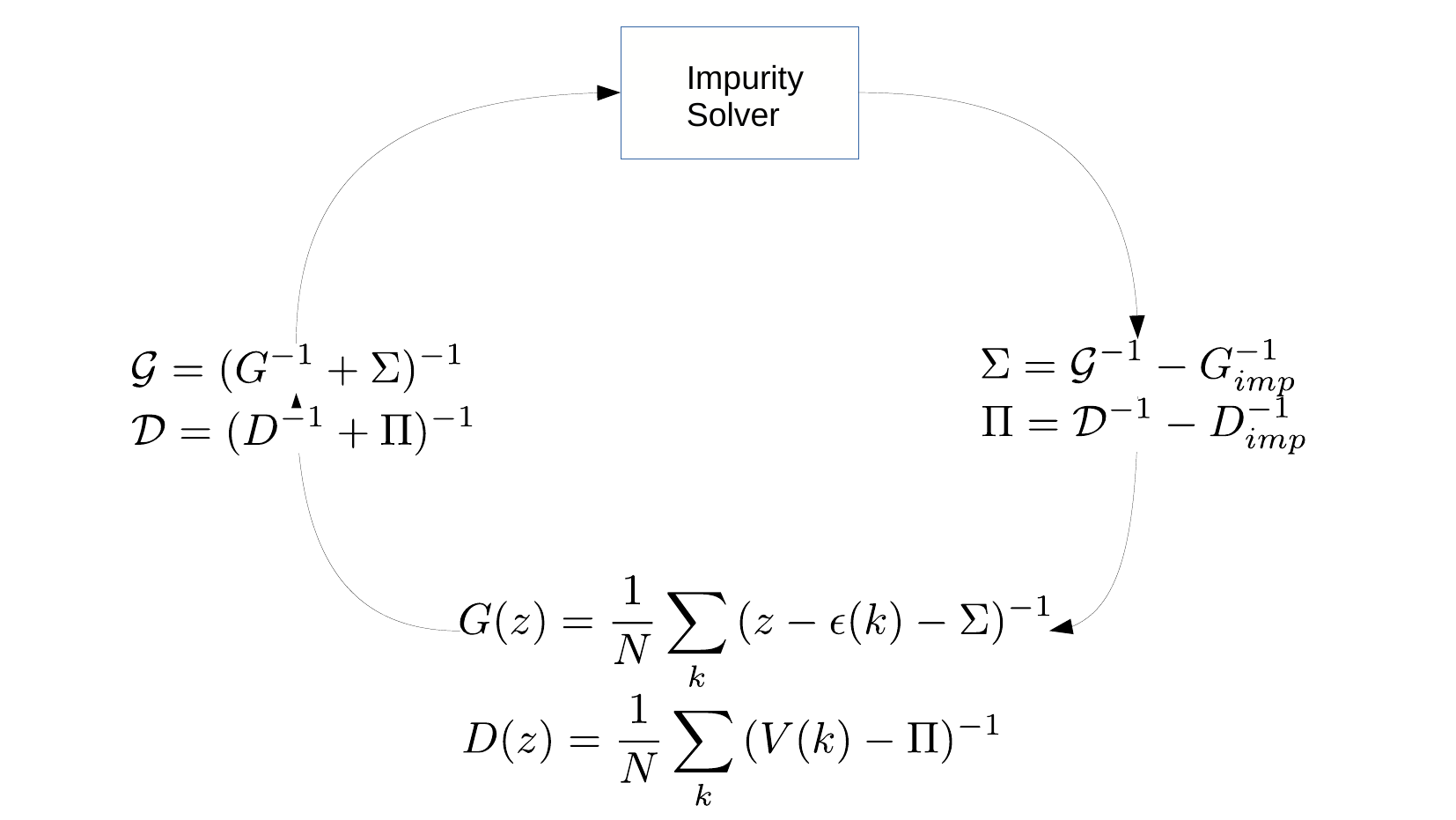}
\caption{EDMFT algorithm: {\bf 1.} The impurity solver provides the impurity Green function for the electrons, $G_{imp}$ and the impurity Green function for the Bosonic field, $D_{imp}$. {\bf 2.} Dyson equation is used to extract the self-energy $\Sigma$ and the polarization, $\Pi$. {\bf 3.} The self-energy and the polarization is coarse-grained over the entire first Brillouin zone to obtain the lattice Green functions. {\bf 4.} The bath Green function for the electrons and the Bosonic field of the impurity problem is feed into the impurity solver. {\bf 5.} Repeat from step 1 until convergence is obtained for the self-energy and polarization or Green functions.
} 
\label{fig:EDMFT_algorithm}
\end{figure}

The algorithm is thus similar to that of the DMFT, except that at each iteration both the mean fields for the fermionic and for the bosonic baths have to be updated. The algorithm can be summarized as in Fig. \ref{fig:EDMFT_algorithm}

\section{$1/d$ Expansion}
\label{sec:1-over-d}

One of the earliest attempts to consider the effect of finite dimension and move away from the infinite dimension limit is the systematic expansion with respect to the hopping amplitude. This approach has been studied in other contexts, especially in the closely related CPA for disordered systems. \cite{Gebhard_1990,Vlaming_Vollhardt_1992,Uhrig_1996,Janis_Vollhardt_2001,vanDongen_1994} Since the hopping term carries the factor of $1/\sqrt{d}$, an expansion with respect to the hopping regains the dimensional dependence at the expense of including multiple sites. 
Work by Schiller and Ingersent studied the case of two impurities for the Falicov-Kimball model. \cite{Schiller_Ingersent_1995} 
The approach has also been applied to the half-filled Hubbard model. \cite{Zarand_etal_2000}

The $1/d$ corrections can be recovered by considering a multi-impurity problem. For a single impurity ($n=1$)  or a two-impurity  $(n=2)$ problem,
\begin{eqnarray}
S^{(n)} &&=  \sum_\sigma \sum_{\alpha,\beta = 1}^n 
\int d\tau \int d\tau' {\psi}^{*}_{\alpha \sigma}(\tau) 
({\cal G}^{(n)}_{\alpha,\beta})^{-1} (\tau-\tau') \psi_{\beta \sigma}(\tau') \nonumber \\
&& + \sum_{\alpha = 1}^n U \int d\tau n_{\alpha \uparrow}(\tau) 
n_{\alpha \downarrow}(\tau)\;,
\end{eqnarray}
where $\alpha$ and $\beta$ label the sites for $n=2$. \cite{Schiller_Ingersent_1995,Zarand_etal_2000}
The mean fields ${\cal G}^{(1)}$ and ${\cal G}^{(2)}$ are
 chosen in such a way that the impurity Green functions $G^{(1)}$ and $G^{(2)}$ coincide with the full on-site and nearest neighbor lattice propagators, $G^{\rm lattice}_{00}$ and $G^{\rm lattice}_{01}$:
\begin{equation}
G^{(1)} = G^{(2)}_{11} =  G^{\rm lattice}_{00}\; \phantom{n},
 \phantom{nnn} G^{(2)}_{12} = G^{\rm lattice}_{01}\;.
\end{equation}
For skeleton diagrams (diagrams without self-energy dressing or vertex correction) of  order ${\cal O}(1/d)$, 
the impurity  self energies $\Sigma^{(1)}$ and $\Sigma^{(2)}_{\alpha\beta}$ 
and the diagonal and off-diagonal lattice self energies, 
$\Sigma^{\rm lattice}_0$ and $\Sigma^{\rm lattice}_1$  are related by 
\cite{Schiller_Ingersent_1995,Zarand_etal_2000}
\begin{eqnarray}
\Sigma^{\rm lattice}_0 &=& \Sigma^{(1)} + 2d(\Sigma^{(2)}_{11} - \Sigma^{(1)})\;,
\label{eq:Sigma0} \\
\Sigma^{\rm lattice}_1 &=& \Sigma^{(2)}_{12}\;.\label{eq:Sigma12}
\end{eqnarray}
The lattice Green function can be obtained from the lattice self-energy:
\begin{equation} 
G_{lm}^{\rm lattice}(i\omega) = {1\over1 +\sqrt{d} \; \Sigma^{\rm lattice}_1}
G^{0}_{lm}\left({i\omega - \Sigma^{\rm lattice}_0(i\omega) \over
1 +\sqrt{d} \; \Sigma^{\rm lattice}_1(i\omega)}\right)\;,
\end{equation}
where $G^{0}_{lm}(z)$ and $G_{lm}^{\rm lattice}(z)$ denote the unperturbed 
and dressed lattice propagators between sites $l$ and $m$, respectively.
\cite{Schiller_Ingersent_1995,Zarand_etal_2000}

Among the shortcomings of the $1/d$ expansion are its limitations in the description of long range fluctuations. Perhaps more importantly, the truncation at finite order may lead to non-analytic properties of some dynamical quantities. \cite{Pruschke_etal_2001} The method is an example of the nested cluster scheme and is also related to the recently proposed self-energy embedding theory. \cite{Biroli_2004,Kananenka_etal_2015,Zgid_Gull_2017}

\section{Multiscale many body} 
\label{sec:MSMB}

Given the challenge of solving a large cluster by numerical methods  and the lack of accurate analytical approaches, perturbation theory is a possible route for the exploration of physics beyond local theories. Initial efforts aimed to use perturbative methods as solvers for the cluster impurity problem. Notably, the fluctuation exchange method was used as an impurity cluster solver. \cite{k_aryanpour_03a} The fluctuation-exchange or FLEX is one of the simplest methods to incorporate correlations among different channels. It is, therefore, a conceptually appealing approach for systems in which particle-hole or particle-particle vertex fluctuations are not small.
If correlations of the two particle fluctuations are ignored, one obtains the second order perturbation theory for which the irreducible vertex is replaced by the bare vertex. 
FLEX allows vertex contributions from different channels to be correlated, thus the name of fluctuation exchange approximation.

The general scheme of the multiscale many body method as envisioned by Jarrell and collaborators is to construct a theory which allows for the treatment of different length scales by different approaches. \cite{m_jarrell_07,c_slezak_06b} The short length scale is addressed by some highly accurate numerical approach, such as quantum Monte Carlo. The long length scale is treated at a mean field level, while the intermediate length scale is treated by some form of perturbative technique. 

An early proposition was to supplement the Quantum Monte Carlo calculation on small cluster sizes with FLEX on larger clusters via a self-energy self-consistency scheme between the two methods. \cite{j_hague_04}
It was generalized in 2006 to address short and long length scales within the DCA formalism while the intermediate length scale would be treated by the parquet formalism. \cite{c_slezak_06b} In this approach, the connection between the two methods is established by the fully irreducible vertex from the DCA calculation on the cluster. The promise of the approach stems from the better scaling of the parquet formalism compared to that of QMC as a cluster solver.

\begin{figure}[htbp]
\includegraphics*[width=10.0cm, height=6.0cm]{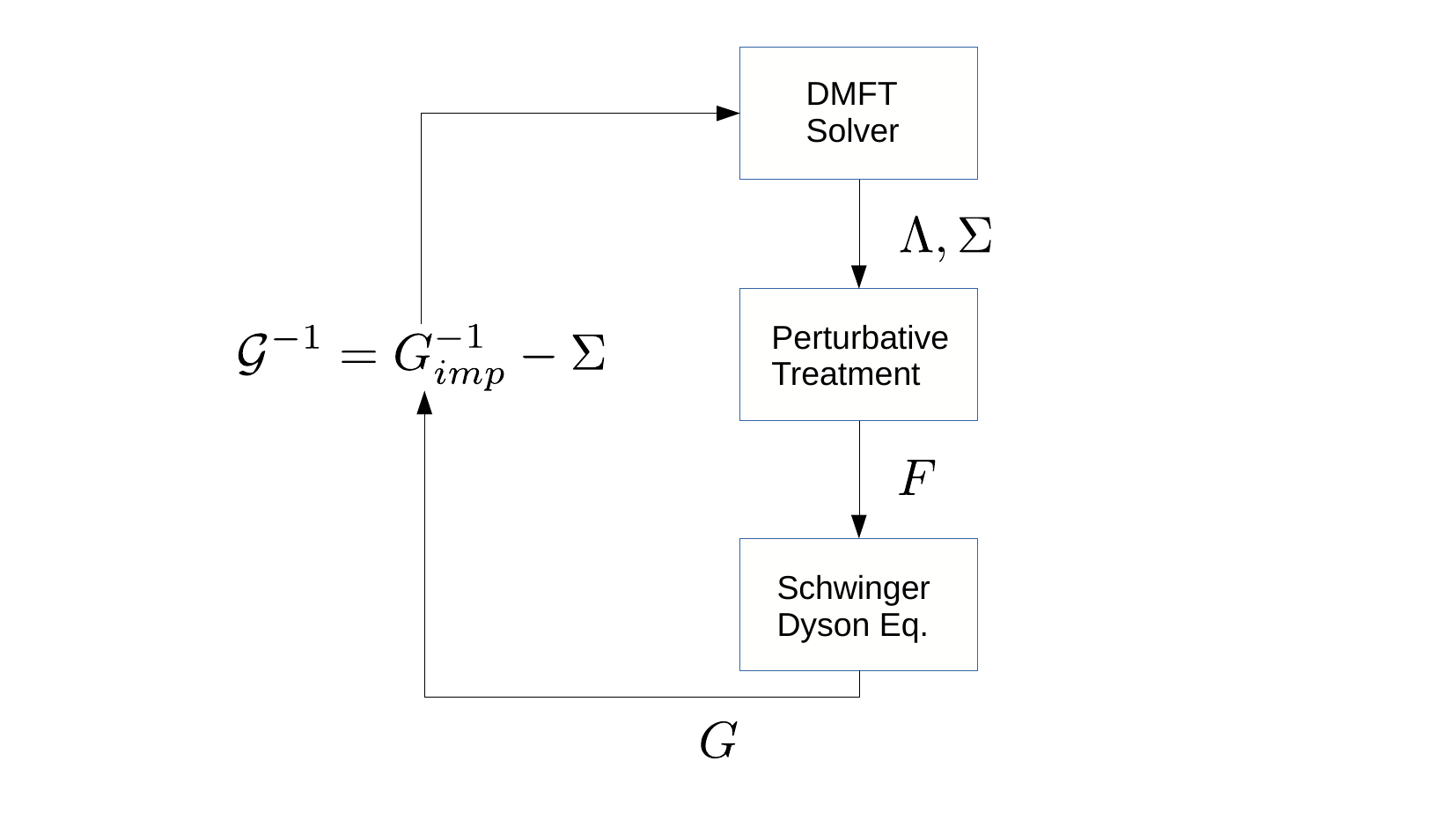}
\caption{Algorithm for the
multiscale many body method. {\bf  1.} The fully irreducible vertex ($\Lambda$) and optionally the self-energy ($\Sigma$) are obtained from the DMFT or DCA solver. {\bf 2.} The parquet method or some other perturbative method is used to calculate the full vertex to restore spatial dependence or enhance spatial resolutions in the full vertex ($F$). {\bf 3.} The full vertex is fed into Schwinger Dyson equation and Dyson equation to update the self-energy and the Green function respectively. {\bf 4.} The Green function can be coarse-grained to be fed back to the DMFT or DCA solver as the bath Green function for self-consistent calculations. } 
\label{fig:msmb_algorithm}
\end{figure}

The general construction of the multiscale many body method, as summarized in Fig.~\ref{fig:msmb_algorithm}, is obviously rather general. There is in fact plenty of freedom in the choice of a solver for the intermediate length scale. Indeed, this area of research has been the subject of significant activity over the past decade or so. Nevertheless, most of the developed methods are based on some simplification of the original proposal, either in picking a certain subclass of diagrams from the parquet formalism or in simplifying the solution by numerical techniques.

\begin{figure}[htbp]
\label{fig:msmb_pic}
\includegraphics*[width=9.0cm, height=4.0cm, trim=0.5cm 6cm 0.5cm 4cm ]{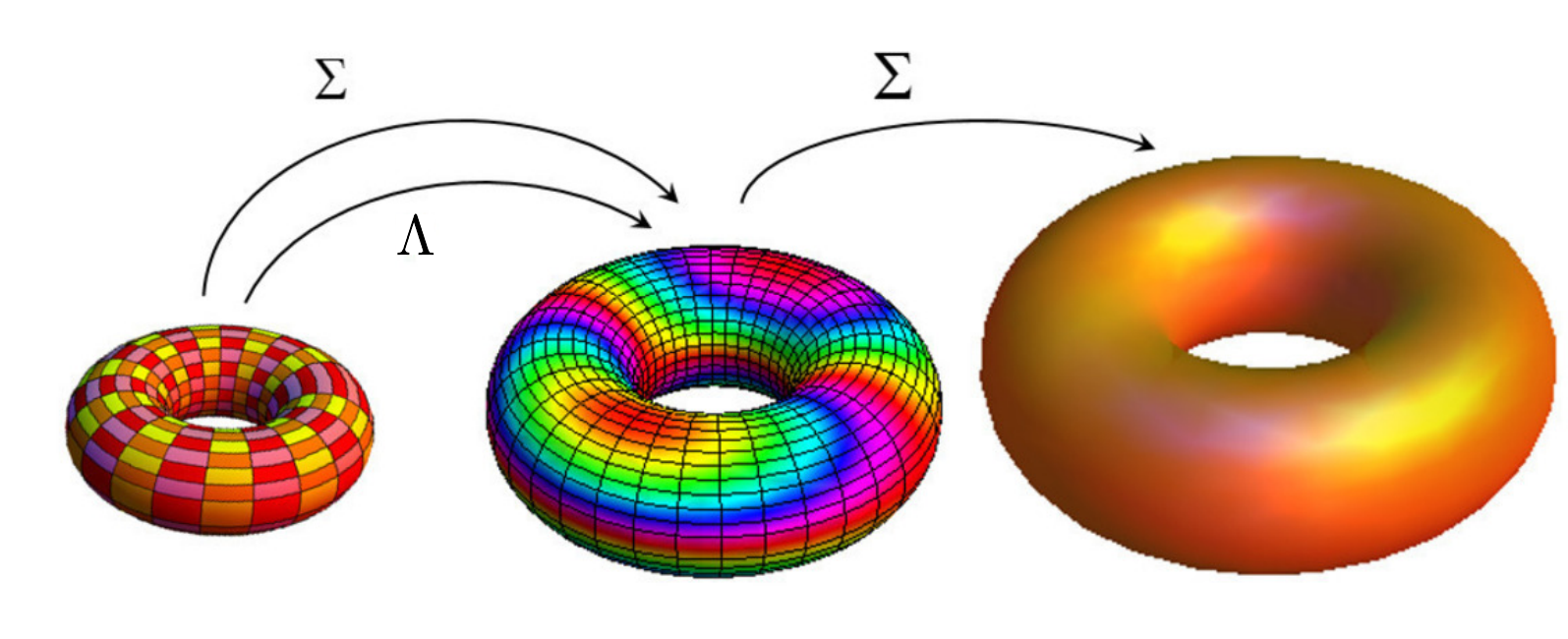}
\caption{Conceptual idea of the multiscale many body approach. 
The original lattice model is mapped onto a small cluster (yellow-orange-red colors), embedded
in a larger cluster (rainbow colors) embedded in a mean field (orange color). The information passed between the clusters and the effective medium is composed of irreducible quantities like the self-energy and the fully irreducible vertex function
} 
\end{figure}

\section{Diagrammatic methods and the Parquet Formalism}
\label{sec:Parquet}

A major route for perturbative expansions around the DMFT solution is based on the parquet diagrams. This approach encapsulates many of the approximations that have emerged in this field, including the dynamical vertex approximation. For this reason, we review the parquet method in rather self-contained detail in this section. 

Standard diagrammatic perturbative expansions attempt to describe all the scattering processes as single or two-particle Feynman diagrams. In the single-particle formulation the self-energy describes the many-body processes that renormalize the motion of a particle in the interacting background of all the other particles. In the two-particle context, one is able to probe the interactions between particles using the so-called vertex functions, which are matrices describing two particle scattering processes. For example, the reducible (full) two-particle vertex
$F^{ph}(12;34)$ describes the scattering amplitude of a particle-hole pair  from its initial state $\left|3,4\right>$ into the final state $\left|1,2\right>$. Here, $i = 1,2,3,4$ represents a set of indices which combines the momentum $\mathbf{k}_i$, the Matsubara frequency $i\omega_{n_i}$ and, if needed, the spin $\sigma_i$ and band index $m_i$. Since the total momentum and energy of the vertex are conserved, it is convenient to adopt the notation $F^{ph}(2-4)_{1,3}$ for the numerical implementation on the single band Hubbard model.  Other representations are also possible. \cite{Karrasch_etal_2008} 

In general, depending on how particles or holes are involved in the scattering processes, one can define three different two-particle scattering channels. These are the particle-hole (p-h) horizontal channel, the p-h vertical channel and the particle-particle (p-p) channel. The parquet formalism is, in essence, a method for summing up diagrams that characterize scattering processes at the two-particle level. From another perspective, the diagrams are generated by inserting the one loop, second order, diagrams repeatedly into itself. Without channel mixing, this is equivalent to the random phase approximation. With the mixing of three channels, this becomes the parquet formalism.

\begin{figure}[htbp]
\includegraphics*[width=8.0cm, height=6.0cm]{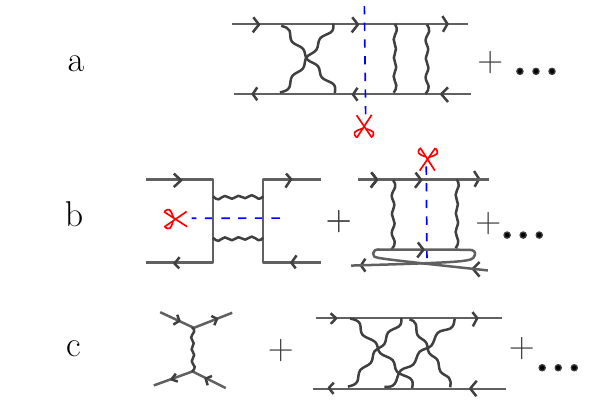} 
\caption{Illustration of the reducibility of a diagram in the particle-hole horizontal channel. (a) The diagram is reducible in the particle-hole channel in the sense that the diagram can be separated by cutting two horizontal Green function lines. (b) Examples of diagrams which are irreducible with respect to the particle-hole horizontal channel. The one on the left hand side is reducible in the particle-hole vertical channel. The one on the right hand side is reducible in the particle-particle  channel. (c) Examples of fully irreducible diagrams. They cannot be separated into two parts by cutting two Green functions lines in any one of the three channels. From Ref.~[\onlinecite{Yang_etal_2009}].
} 
\label{fig:reducibility}
\end{figure}

The vertices can be categorized by extending the notion of diagram reducibility to the two-particle level as illustrated by Fig.~\ref{fig:reducibility}. At the one particle level, a diagram is said to be reducible if it can be split in two disconnected parts by breaking a single Green's function line. A two-particle diagram will be said to be irreducible if it can not be separated in two disconnected parts by breaking two Green function lines in the same channel. \cite{n_bickers_98} It will be said to be fully irreducible if it can not be separated in two disconnected parts by breaking two-Green's function lines in any channel. In the single particle formalism, the Green function is related to the self-energy containing all single-particle irreducible diagrams by the Dyson equation. A connection is made between the single-particle and the two-particle diagrams by the Schwinger-Dyson equation that connects the self-energy to the (full/reducible) vertex  $F$ containing all allowed diagrams in a given channel. The subset of all two-level diagrams in the full vertex that are irreducible in the same channel is known as the irreducible vertex $\Gamma$. The subset of irreducible vertices that are irreducible in any channel is called the fully irreducible vertex $\Lambda$.

It is worth noting that the above idea for the decomposition is not unique. One can devise other possible decompositions. For example, a recent attempt is to decompose diagrams in terms of the fermion-boson vertex. We will not explore this direction in detail in this review. The above decomposition is the most natural one in the sense that the method can be easily understood in terms of an iterative process. The higher order diagrams are all generated by iteratively replacing the vertex function at a lower order approximation.

Furthermore, we are mostly interested in models that preserve the $SU(2)$ spin rotation symmetry. Since this symmetry is always obeyed for the two-dimensional calculations at non-zero temperature, it is convenient to preserve this symmetry. This is accomplished by decomposing the vertices in the so-called spin-diagonalized representation. In this representation, the spin degrees of freedom decompose the particle-hole channel into the density and the magnetic channels, and the particle-particle channel into the spin singlet and the spin triplet channels which we denote as $d$-channel, $m$-channel, $s$-channel, and $t$-channel respectively. \cite{n_bickers_98} They are defined for the irreducible vertex as follows,

\begin{eqnarray}
\Gamma_{d} &=& \Gamma^{PH}_{\uparrow \uparrow ; \uparrow \uparrow} + \Gamma^{PH}_{\uparrow \uparrow ; \downarrow \downarrow}, \\
\Gamma_{m} &=& \Gamma^{PH}_{\uparrow \uparrow ; \uparrow \uparrow} - \Gamma^{PH}_{\uparrow \uparrow ; \downarrow \downarrow}, \\
\Gamma_{s} &=& \Gamma^{PP}_{\uparrow \downarrow ; \uparrow \downarrow} - \Gamma^{PP}_{\uparrow \downarrow ; \downarrow \uparrow}, \\
\Gamma_{t} &=& \Gamma^{PP}_{\uparrow \downarrow ; \uparrow \downarrow} + \Gamma^{PP}_{\uparrow \downarrow ; \downarrow \uparrow},
\end{eqnarray}
and similarly for $F$ and $\Lambda$.

The formalism is completed by equations that connect the different types of vertices. The full vertex is related to the irreducible vertex by the Bethe-Salpeter equation and the irreducible vertex is in turn related to the fully irreducible vertex by the parquet equation. We reproduce the full set of equations for the parquet formulation in the spin diagonalized representation in the following. 

The Schwinger-Dyson equation that connects the vertex to the self-energy is
\begin{widetext}
\begin{eqnarray}
\Sigma(P) &=& -\frac{UT^2}{4N} \sum_{P^\prime,Q} \{G(P^\prime) G(P^\prime+Q)
G(P-Q) (F_d(Q)_{P-Q,P^\prime}-F_m(Q)_{P-Q,P^\prime}){\nonumber} \\
&& \;\;\;\;\;\;\;\;\;\;\;\;\;\; + G(-P^\prime) G(P^\prime+Q) G(-P+Q)
(F_s(Q)_{P-Q,P^\prime}+F_t(Q)_{P-Q,P^\prime})\},
\end{eqnarray}
\end{widetext}
where $G$ is the single-particle Green function, which itself can be
calculated from the self-energy using the Dyson equation,
\begin{eqnarray}
G^{-1}(P) &=& G_0^{-1}(P)\; - \; \Sigma(P),
\end{eqnarray}
where $G_{0}$ is the bare Green function. Here, the indices $P$, $P^\prime$ and $Q$ combine momentum ${\bf k}$
and Matsubara frequency $i\omega_n$, i.e.\ $P=({\bf k},i\omega_n)$.

The reducible and the irreducible vertices in a given channel are
related by the Bethe-Salpeter equation,
\begin{equation}
F_{r}(Q)_{P,P^\prime} = \Gamma_{r}(Q)_{P,P^\prime} +
\Phi_{r}(Q)_{P,P^\prime}, 
\label{SD_BSPH_EQ}
\end{equation}
\begin{equation}
F_{r^\prime}(Q)_{P,P^\prime} = \Gamma_{r^\prime}(Q)_{P,P^\prime} +
\Psi_{r^\prime}(Q)_{P,P^\prime},
\label{SD_BSPP_EQ}
\end{equation}
where $r=d\;\mbox{or}\;m$ for the density and magnetic channels
and $r^\prime=s\;\mbox{or}\;t$ for the spin singlet and
spin triplet channels. The vertex ladders are
defined as
\begin{eqnarray}
\Phi_{r}(Q)_{P,P^\prime} \equiv 
\;\;\;\;\;\;\;\;\;\;\;\;\;\;\;\;\;\;\;\;\;\;\;\;\;\;\;\;\;\;\;\;\;\;\;\;\;\;\;\;\;\;\;\;\;
\label{eq:Phi_ladder} \\ \nonumber
\sum_{P^{\prime\prime}}F_{r}(Q)_{P,P^{\prime\prime}}\chi_0^{ph}(Q)_{P^{%
\prime\prime}} \Gamma_{r}(Q)_{P^{\prime\prime},P^\prime}, \\
\Psi_{r^\prime}(Q)_{P,P^\prime} \equiv 
\;\;\;\;\;\;\;\;\;\;\;\;\;\;\;\;\;\;\;\;\;\;\;\;\;\;\;\;\;\;\;\;\;\;\;\;\;\;\;\;\;\;\;\;\;
\label{eq:Psi_ladder}  \\ \nonumber
\sum_{P^{\prime\prime}}F_{r^\prime}(Q)_{P,P^{\prime\prime}}\chi_0^{pp}(Q)_{P^{%
\prime\prime}} \Gamma_{r^\prime}(Q)_{P^{\prime\prime},P^\prime},
\end{eqnarray}
where $\chi_0$, the bare susceptibility, is the product of two single-particle Green functions.  

The parquet equations in the spin diagonalized representation are 
\begin{widetext}
\begin{eqnarray}
\label{eq:Gamma_d}
\Gamma_d(Q)_{P{P^\prime}} &=& \Lambda_d(Q)_{P{P^\prime}} - {\frac{1 }{2}}%
\Phi_d({P^\prime}-P)_{P,P+Q} - {\frac{3 }{2}}\Phi_m({P^\prime}-P)_{P,P+Q} 
\\
&& \;\;\;\;\;\;\;\;\;\;\;\;\;\;\; + \; {\frac{1 }{2}}\Psi_s(P+{P^\prime}%
+Q)_{-P-Q,-P} + {\frac{3 }{2}}\Psi_t(P+{P^\prime}+Q)_{-P-Q,-P}, \nonumber
\end{eqnarray}
\begin{eqnarray}
\label{eq:Gamma_m}
\Gamma_m(Q)_{P{P^\prime}} &=& \Lambda_m(Q)_{P{P^\prime}} - {\frac{1 }{2}}%
\Phi_d({P^\prime}-P)_{P,P+Q} + {\frac{1 }{2}}\Phi_m({P^\prime}-P)_{P,P+Q} 
\\
&& \;\;\;\;\;\;\;\;\;\;\;\;\;\;\;\; - \; {\frac{1 }{2}}\Psi_s(P+{P^\prime}%
+Q)_{-P-Q,-P} + {\frac{1 }{2}}\Psi_t(P+{P^\prime}+Q)_{-P-Q,-P}, \nonumber
\end{eqnarray}
\begin{eqnarray}
\label{eq:Gamma_s}
\Gamma_s(Q)_{P{P^\prime}} &=& \Lambda_s (Q)_{P{P^\prime}} + {\frac{1 }{2}}%
\Phi_d({P^\prime}-P)_{-{P^\prime},P+Q} - {\frac{3 }{2}}\Phi_m({P^\prime}%
-P)_{-{P^\prime},P+Q}  \\
&& \;\;\;\;\;\;\;\;\;\;\;\;\;\;\; + \; {\frac{1 }{2}}\Phi_d(P+{P^\prime}%
+Q)_{-{P^\prime},-P} - {\frac{3 }{2}}\Phi_m(P+{P^\prime}+Q)_{-{P^\prime},-P}, \nonumber
\end{eqnarray}
\begin{eqnarray}
\label{eq:Gamma_t}
\Gamma_t(Q)_{P{P^\prime}} &=& \Lambda_t (Q)_{P{P^\prime}} + {\frac{1 }{2}}%
\Phi_d({P^\prime}-P)_{-{P^\prime},P+Q} + {\frac{1 }{2}}\Phi_m({P^\prime}%
-P)_{-{P^\prime},P+Q}  \\
&& \;\;\;\;\;\;\;\;\;\;\;\;\;\;\; - \; {\frac{1 }{2}}\Phi_d(P+{P^\prime}%
+Q)_{-{P^\prime},-P} - {\frac{1 }{2}}\Phi_m(P+{P^\prime}+Q)_{-{P^\prime},-P}. \nonumber
\end{eqnarray}
\end{widetext}

It is important to note that if we substitute the
irreducible vertices $\Gamma$ (Eqs.~\ref{eq:Gamma_d}, \ref{eq:Gamma_m}, \ref{eq:Gamma_s}, and \ref{eq:Gamma_t}) into the Bethe-Salpeter equation (Eqs.~\ref{SD_BSPH_EQ} and \ref{SD_BSPP_EQ}) the crossing symmetries (symmetry relations of the vertex that are a consequence of the Pauli exclusion principle for identical fermionic particles) in the full vertex $F$ is automatically satisfied regardless of the numerical values of the vertex ladders $\Phi$ and $\Psi$, assuming the fully irreducible vertices, $\Lambda$, obey the crossing symmetries. We write all the full vertices explicitly in the following, using only the vertex ladders, $\Phi$, $\Psi$, and the fully irreducible vertices, $\Lambda$. 
\begin{widetext}

\begin{eqnarray}
\label{eq:F_d}
F_{d}(Q)_{P,P^\prime} = \Lambda_d(Q)_{P{P^\prime}} - {\frac{1 }{2}} \Phi_d({P^\prime}-P)_{P,P+Q} - {\frac{3 }{2}}\Phi_m({P^\prime}-P)_{P,P+Q} \\ 
+ {\frac{1 }{2}}\Psi_s(P+{P^\prime}+Q)_{-P-Q,-P} + {\frac{3 }{2}}\Psi_t(P+{P^\prime}+Q)_{-P-Q,-P} + \Phi_{d}(Q)_{P,P^\prime}; \nonumber
\end{eqnarray}

\begin{eqnarray}
\label{eq:F_m}
F_{m}(Q)_{P,P^\prime} &=& \Lambda_m(Q)_{P{P^\prime}} - {\frac{1 }{2}}\Phi_d({P^\prime}-P)_{P,P+Q} + {\frac{1 }{2}}\Phi_m({P^\prime}-P)_{P,P+Q} \\ 
&-&  {\frac{1 }{2}}\Psi_s(P+{P^\prime}+Q)_{-P-Q,-P} + {\frac{1 }{2}}\Psi_t(P+{P^\prime}+Q)_{-P-Q,-P} + \Phi_{m}(Q)_{P,P^\prime}; \nonumber
\end{eqnarray}

\begin{eqnarray}
\label{eq:F_s}
F_{s}(Q)_{P,P^\prime} &=& \Lambda_s (Q)_{P{P^\prime}} + {\frac{1 }{2}} \Phi_d({P^\prime}-P)_{-{P^\prime},P+Q} - {\frac{3 }{2}}\Phi_m({P^\prime}-P)_{-{P^\prime},P+Q}  \\ 
&+&  {\frac{1 }{2}}\Phi_d(P+{P^\prime}+Q)_{-{P^\prime},-P} - {\frac{3 }{2}}\Phi_m(P+{P^\prime}+Q)_{-{P^\prime},-P} + \Psi_{s}(Q)_{P,P^\prime}; \nonumber
\end{eqnarray}

\begin{eqnarray}
\label{eq:F_t}
F_{t}(Q)_{P,P^\prime} &=& \Lambda_t (Q)_{P{P^\prime}} + {\frac{1 }{2}} \Phi_d({P^\prime}-P)_{-{P^\prime},P+Q} + {\frac{1 }{2}}\Phi_m({P^\prime}-P)_{-{P^\prime},P+Q} \\  
&-& {\frac{1 }{2}}\Phi_d(P+{P^\prime}+Q)_{-{P^\prime},-P} - {\frac{1 }{2}}\Phi_m(P+{P^\prime}+Q)_{-{P^\prime},-P} + \Psi_{t}(Q)_{P,P^\prime}. \nonumber
\end{eqnarray}

\end{widetext}

\begin{figure}[htbp]
\includegraphics*[width=10.cm, height=5.2cm]{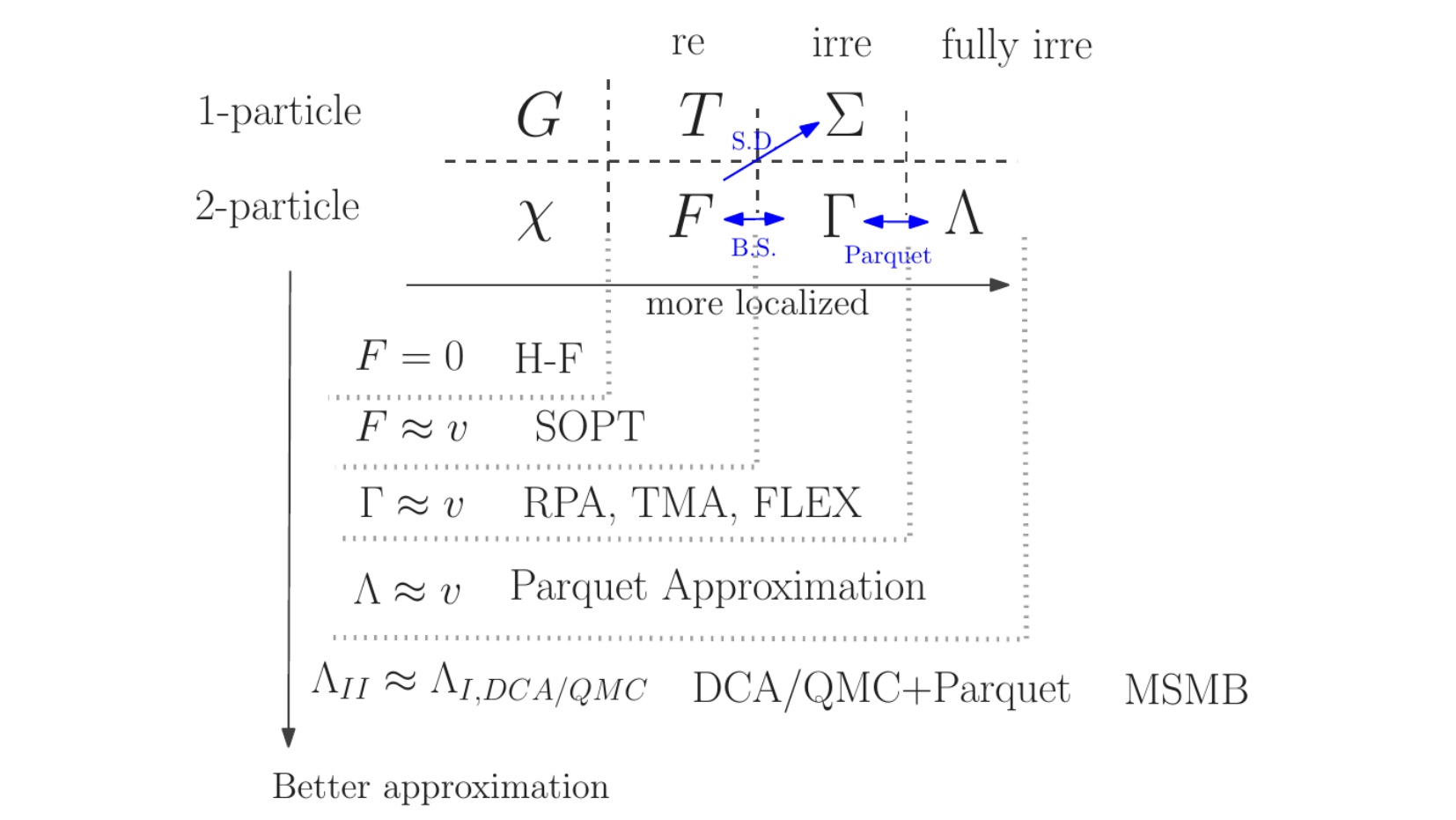} 
\caption{Hierarchy of approximations encompassed by the parquet approximation. See the text for details. S.D. and B.S. stand for Schwinger-Dyson equation and Bethe-Salpeter equation respectively. $v$ is the bare vertex characterizing the lowest order two-particle scattering processes.
} 
\label{fig:MSMB}
\end{figure}

The full parquet formalism encompasses a variety of approximations that are widely used in condensed matter physics and materials science. The hierarchy of these different approximations  is neatly summarized in  Fig. \ref{fig:MSMB}.

\begin{itemize}
 
\item Hartree-Fock and second order perturbation theory: At the highest level, we might make the approximation on the two-particle Green function (analogous to the conventional single particle Hartree-Fock (HF) perturbation) such that the four-point correlation function can be factorized as a product of two two-point correlation functions. It is equivalent to ignoring the contribution from the full vertex functions. From the Schwinger-Dyson equation, the self-energy has two contributions a Hartree-Fock term and a second order perturbation theory term. 

\item Self-consistent second order perturbation theory: Substituting the bare vertex for the full vertex in the Schwinger-Dyson equation and solving for the self-energy self-consistently results in the self-consistent second order perturbation theory. 

\item Random-Phase Approximation (RPA): The irreducible vertex in the longitudinal charge channel is approximated by the bare Coulomb interaction. And then the Bethe-Salpeter equation is used to sum up over all the ring-type diagrams. \cite{Bohm_Pines_1951,Bohm_Pines_1952,Bohm_Pines_1952}

\item T-Matrix Approximation (TMA): Similar to RPA, the irreducible vertex in the
transverse particle-hole channel or particle-particle channel, instead of longitudinal particle-hole channel, is approximated by the bare Coulomb interaction. And then the Bethe-Salpeter equation is used to sum all the ladder-type (instead of ring-type in RPA) diagrams. \cite{Bethe_Goldstone_1957}

\item Fluctuation Exchange Approximation (FLEX): A combination of RPA and T-matrix approximation, such that the fluctuations in different channels are treated equally. \cite{n_bickers_89}

\end{itemize}

The parquet formalism dates back to the 50's, but as can be seen from the analytical form of the governing equations, a general analytical solution is rather hard to track. \cite{Pomeranchuk_etal_1956, Dominicis_64} Over the years various approximations to simplify the equations have been proposed to solve a set of problems ranging from the Anderson impurity to nuclear structure.
\cite{Roulet_etal_1969,Yakovenko_1993,Brazovskii_1972a,Kleinert_Schlegel_1995,n_bickers_92,Janis_Augustinsky_2007,Janis_Augustinsky_2008,Augustinsky_Janis_2011,Janis_2001,Janis_2009,n_bickers_98,n_bickers_91,d_hess_96,j_luo_93,v_janis_99,Kusunose_2010,Jackson_etal_1985,Jackson_Smith_1987,Pfitzner_Wolfle_1987,Weiner_1970,Weiner_1971,Yeo_Moore_1996a,Yeo_Moore_1996b,Yeo_Moore_2001,Yeo_etal_2006,Shishanin_Ziyatdinov_2003,Arefeva_Zubarev_1996,Bergli_Hjorth-Jensen_2011,Janis_etal_2019}
In addition, a numerical solution is computationally demanding. This difficulty is in general common to theories that involve the two-particle vertex functions. The computational difficulty arises mostly from the memory requirements to store the vertex functions as they are four-legged objects unlike the single particle quantities that only have two legs. If momentum and energy conservation are implemented, the single particle quantities scale linearly with the size of the space-time grid, on the other hand the vertex function scales as the third power of the size of the  space-time grid. This challenge is not insurmountable and may be overcome with appropriate parallelization. In fact, the advent of petascale computing enabled the first solution for the two-dimensional problem.\cite{Yang_etal_2009,k_tam_13}

\begin{figure}[htbp]
\includegraphics*[width=10.0cm, height=6.0cm]{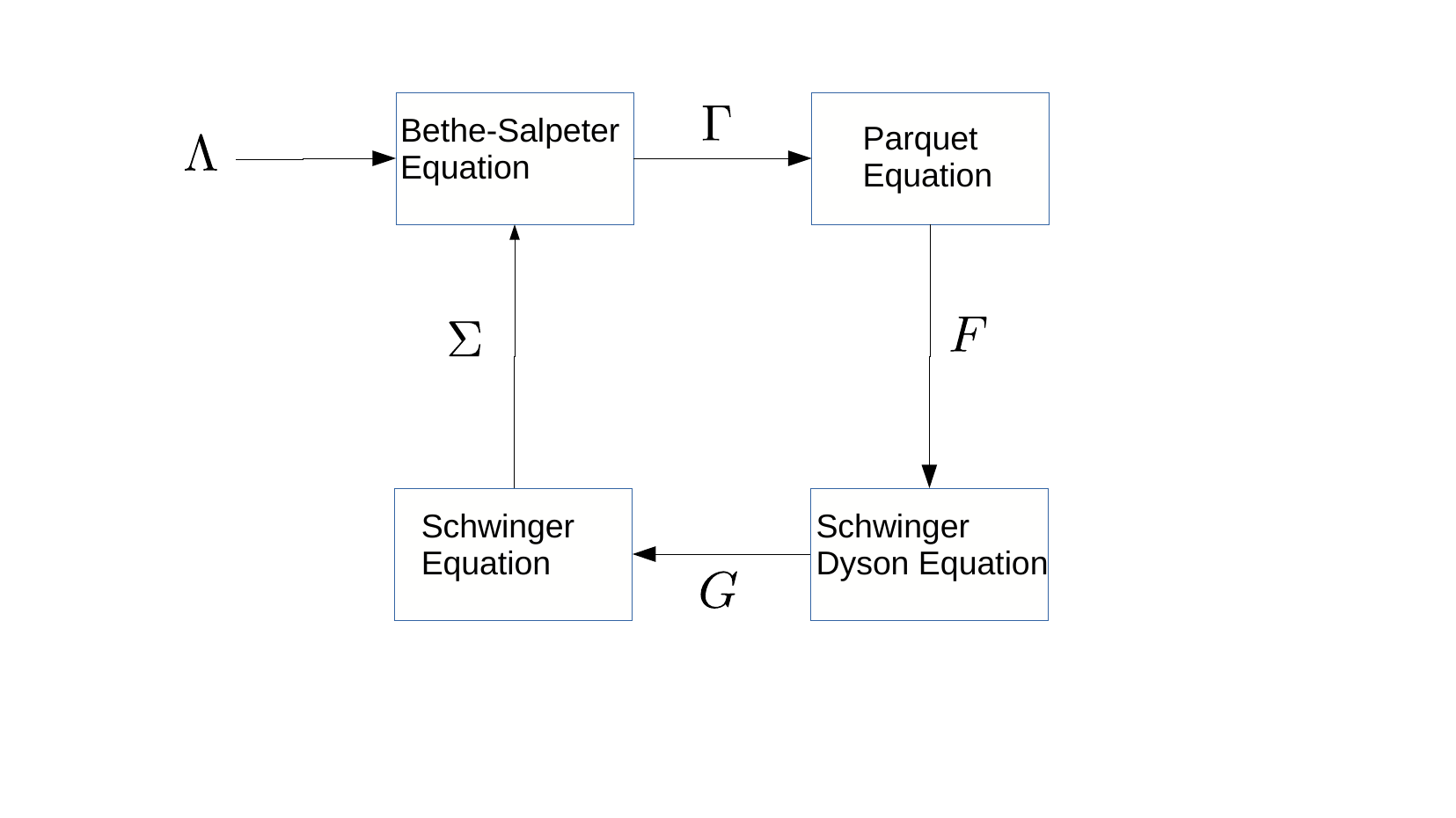}
\caption{
Algorithm for the
parquet method. {\bf 1.} Read in the fully irreducible vertex, $\Lambda$. {\bf 2.} Solve the Bethe Salpeter equation for the irreducible vertex, $\Gamma$. {\bf 3.} Solve the parquet equations for the full vertex, $F$. {\bf 4.} Use the Dyson Schwinger equation to solver for the self-energy. {\bf 5.} The Green function is updated by the Schwinger equation. Repeat from  step 2 until convergence in the full vertex and self-energy is obtained.
} 
\label{fig:parquet_algorithm}
\end{figure}

The algorithm for the numerical solution of the parquet formalism is summarized in Fig.~\ref{fig:parquet_algorithm}.


Perhaps the most challenging problem from the numerical point of view is the difference in the nature of the single particle and the two-particle functions. The single particle quantities in the Matsubara frequency space do not diverge. On the other hand, the two particle vertex has strong divergences when the metallic phase is unstable at the momentum and energy corresponding to an instability. For instance, in the Hubbard model, the particle-hole vertex at the verge of the antiferromagnetic instability has a strong divergence for momentum transfer $(\pi,\pi)$. For this reason, the vertex is represented by numbers spanning a wide range of values over many orders of magnitude. This clearly is a recipe for possible numerical instabilities in the iterative solution for the vertex functions at low temperature and on the verge of long range order phase transitions. As we will discuss in the rest of this review, various methods have been proposed to improve the stability of the numerical solutions. These include simplifying the equations, or abandoning the self-consistent approach for the vertex functions.

\section{Dynamical vertex approximation (D$\Gamma$A)}
\label{sec:DGammaA}

The original dynamical vertex approximation (D$\Gamma$A) is a simplification of the multiscale many-body method with a pragmatic mindset that limits the calculation to the local fully irreducible vertex using perturbation theory. \cite{Held_etal_2008,a_toschi_07} As discussed in section \ref{sec:Parquet} above, a full parquet solution with space-time resolution is very challenging. A natural scheme to sidestep the difficulty is, in the spirit of DMFT, to consider the vertex to be only time or frequency dependent. The local fully irreducible vertex from a DMFT calculation can then be used as input for the parquet formalism. The method subsequently follows the procedure discussed above for the parquet method, therefore, this aspect will not be repeated in this section.

Since this approach is relatively transparent and numerically practical with modest computational costs, \cite{Held_2014} besides the two-dimensional Hubbard model, it has been applied to several problems including the three-dimensional Hubbard model, the attractive Hubbard model, \cite{DelRe_2019} nanoscopic quantum junction systems, \cite{Valli_etal_2010} and it has also been recently combined with ab initio calculations. \cite{Galler_etal_2017} A more elaborate calculation based on parquet method has been performed for the one-dimensional Hubbard model \cite{Schafer_etal_2017}.

There are a few variations of the dynamical vertex approximation. The full parquet formalism can be solved self-consistently with the fully irreducible vertex from the DMFT solution used as input. One can also consider self-consistency at the level of both the parquet equations and the DMFT equations: this involves finding the fully irreducible vertex from the DMFT solution, then using this to solve the parquet equations. The parquet formalism provides both the full vertex and the dressed Green function. The dressed Green function can in turn be treated as input for the DMFT equation to obtain self-consistency in both the DMFT loop and the parquet equations loop.

Initial applications on the half-filled Hubbard model motivated a further simplification by decoupling the particle-hole channels from the particle-particle channels. 
This simplification can be justified by the physics of the systems of interest. For example, in the Hubbard model near half-filling, the density wave is driven by the nesting in the particle-hole channels and one can argue for choosing the particle-hole ladder summations. On the other hand, when the system is driven by s-wave pairing such as that in the attractive Hubbard model, one can keep only the particle-particle channel. There is no systematic universal argument on which channel should be dominant. For most interesting regimes, such as the d-wave pairing in the Hubbard model, presumably all channels could contribute and one may have no choice but to try to tackle the full set of equations of the parquet formalism. 

\begin{figure}[htbp]
\includegraphics*[width=10.0cm, height=6.0cm]{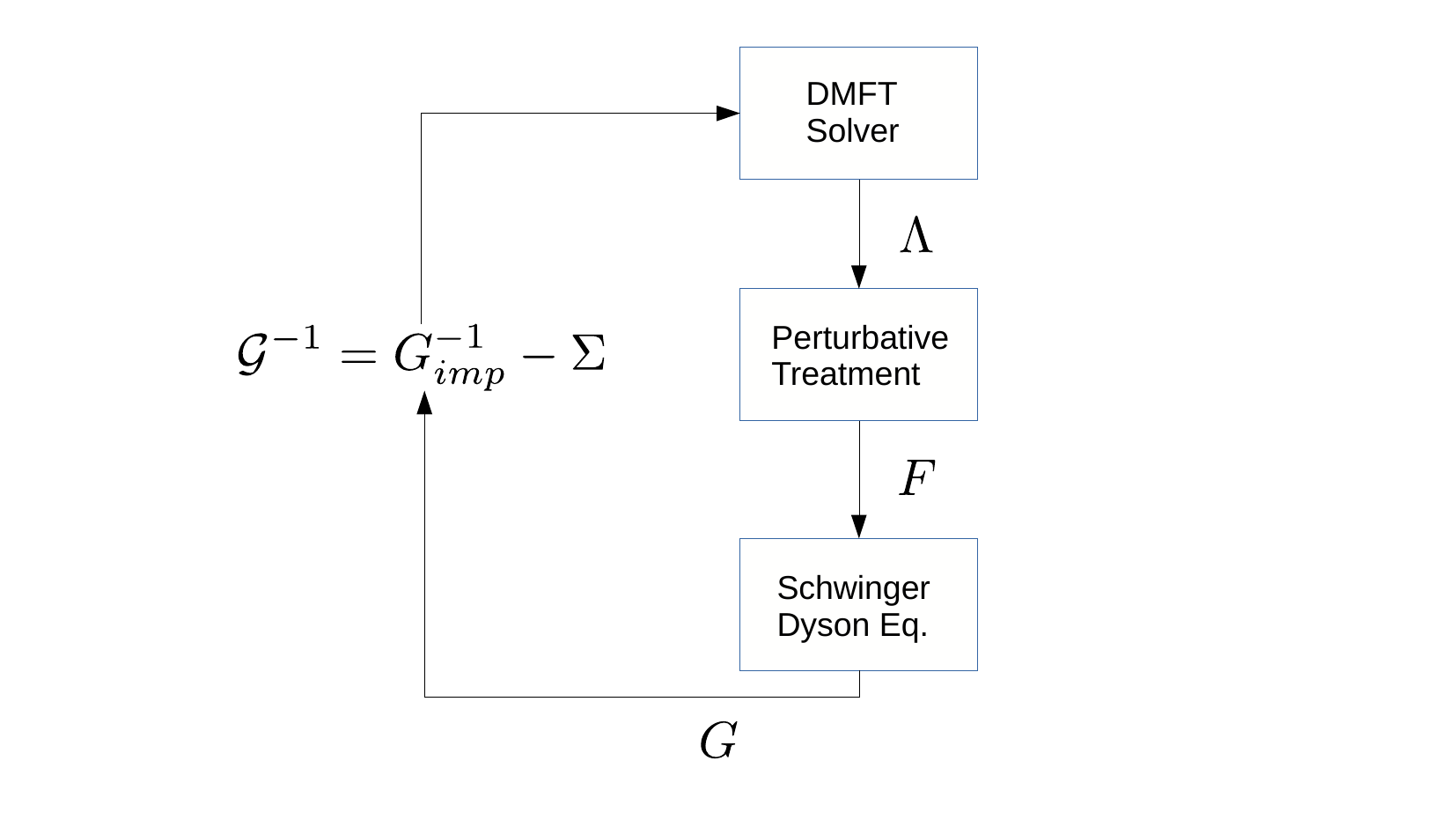}
\caption{ Algorithm for the dynamical vertex approximation. {\bf 1.} The fully irreducible vertex is obtained from the DMFT solver. {\bf 2.} Perturbative methods or the parquet method is used to obtain the full vertex, $F$. {\bf 3.} The Schwinger-Dyson 
equation is solved for the Green function. The results can be read out after this step. For self-consistent calculation, {\bf 4.} The Green function can be coarse-grained to obtain the bath Green function for the impurity problem and the procedure repeated from step 1 until convergence in the Green function is obtained. } 
\label{fig:dga_algorithm}
\end{figure}

The algorithm is summarized in Fig. \ref{fig:dga_algorithm}.

\section{Dual Fermions}
\label{sec:DualFermions}

Another path to the multiscale treatment of correlations in a fermionic system is that of the dual fermions approach that was  built on previous analogous methods for bosonic systems. This approach systematically incorporates nonlocal correlations into the DMFT solution. The method is distinguished from others in that it maps a strongly correlated fermionic lattice onto weakly correlated delocalized fermions. This allows a perturbative treatment of nonlocal correlations using some subsets of allowed diagrams to produce satisfactory corrections on top of the short-length scale correlations that are addressed by an exact solver.

The dual fermion formalism is an extension of the theory by Sarker for strongly correlated system. \cite{Sarker_1988} He proposed a strong coupling expansion of the solution from the atomic limit that predates widespread usage of DMFT. Similar ideas have also been proposed for the study of one dimensional system.  \cite{Pairault_etal_1998} In this theory, the Hubbard model is mapped onto another interacting fermionic model in which the multi-particle hopping-exchange processes appear explicitly.  The formulation is equivalent to the dual fermion formalism as currently known. \cite{a_rubtsov_08,Brener_etal_2008,Li_etal_2008}

Starting from the action of itinerant electrons on a lattice that can be written as:
 \begin{equation}
S\left[ c^*, c \right] = \sum_{\omega, k, \sigma}c^*_{\omega,k,\sigma}\left[i\omega + \mu -h_k \right]c_{\omega,k,\sigma} + \sum_i S_{loc}[c^*, c],
\label{eqn:DF_action}
\end{equation}
where $\mu$ is the chemical potential, $h_k$ is the hoping term in momentum space, $c^*_{\omega, k, \sigma} (c_{\omega, k, \sigma})$ are the Grassmann variables corresponding to the creation (annihilation) operator, and $S_{loc}$ is the local part of the action. The lattice problem can be reframed into that of a set of impurities and an additional term to account for the remaining contributions:
\begin{equation}
S[c^{*},c] = \sum_{i}S_{\textrm{imp}}[c_i^{*},c_i] -\sum_{\omega,k,\sigma}c_{\omega,k,\sigma}^{*}[\Delta(i\omega)-h_{k}
]c_{\omega,k,\sigma}.
\label{eqn:DF_action_imp}
\end{equation}
Revisiting the expression of the partition function, a Hubbard-Stratonovich transformation can be applied on the second term, introducing new fermionic degrees of freedom. The action can then be expressed as:
\begin{eqnarray}
S[c^{*},c;f^{*},f] &=& \sum_{i}S_{restr, i}[c^{*},c;f^{*},f] \nonumber \\
&+& \sum_{\omega, k,\sigma}\frac{f_{\omega, k,\sigma}^{*}\,f_{\omega, k,\sigma}}{g^{2}(i\omega)[\Delta(i\omega)-h_{k}]}.
\label{eqn:ccff1}
\end{eqnarray}

Where $g$ is the single particle DMFT Green function and $S_{restr, i}$ is the action \textit{restricted} to site $i$, and is defined by:
\begin{eqnarray}
S_{restr, i}[c^{*},c;f^{*},f] 
&=& S_{\textrm{imp}}[c^{*},c] \nonumber \\
&+&  \sum_{\omega,\sigma} [f_{\omega,\sigma}^{*}g^{-1}(i\omega)c_{\omega,\sigma}+h.c.].
\label{eqn:ccff2}
\end{eqnarray}
The lattice fermionic degrees of freedom can then be integrated out of the action restricted to the respective sites following:
\begin{eqnarray} 
&\int& \exp \left(-S_{restr, i}\left[c^*, c, f^*, f \right] \right)D\left[c^*, c \right] \nonumber \\
&=& Z_{imp} \exp\left(-\sum_{\omega, \sigma} f^*_{\omega, \sigma}g^{-1}(i\omega)f_{\omega, \sigma} + V \left[ f^*_i, f_i \right] \right).
\end{eqnarray}
This last expression introduces the dual potential $V[f^*_i, f_i]$ in terms of the new fermionic degrees of freedom. It is shown to include all $2n$-vertices  for the impurity with $n=2, 3, 4, ...$.

\begin{figure}[htbp]
\includegraphics*[width=10.0cm, height=6.0cm]{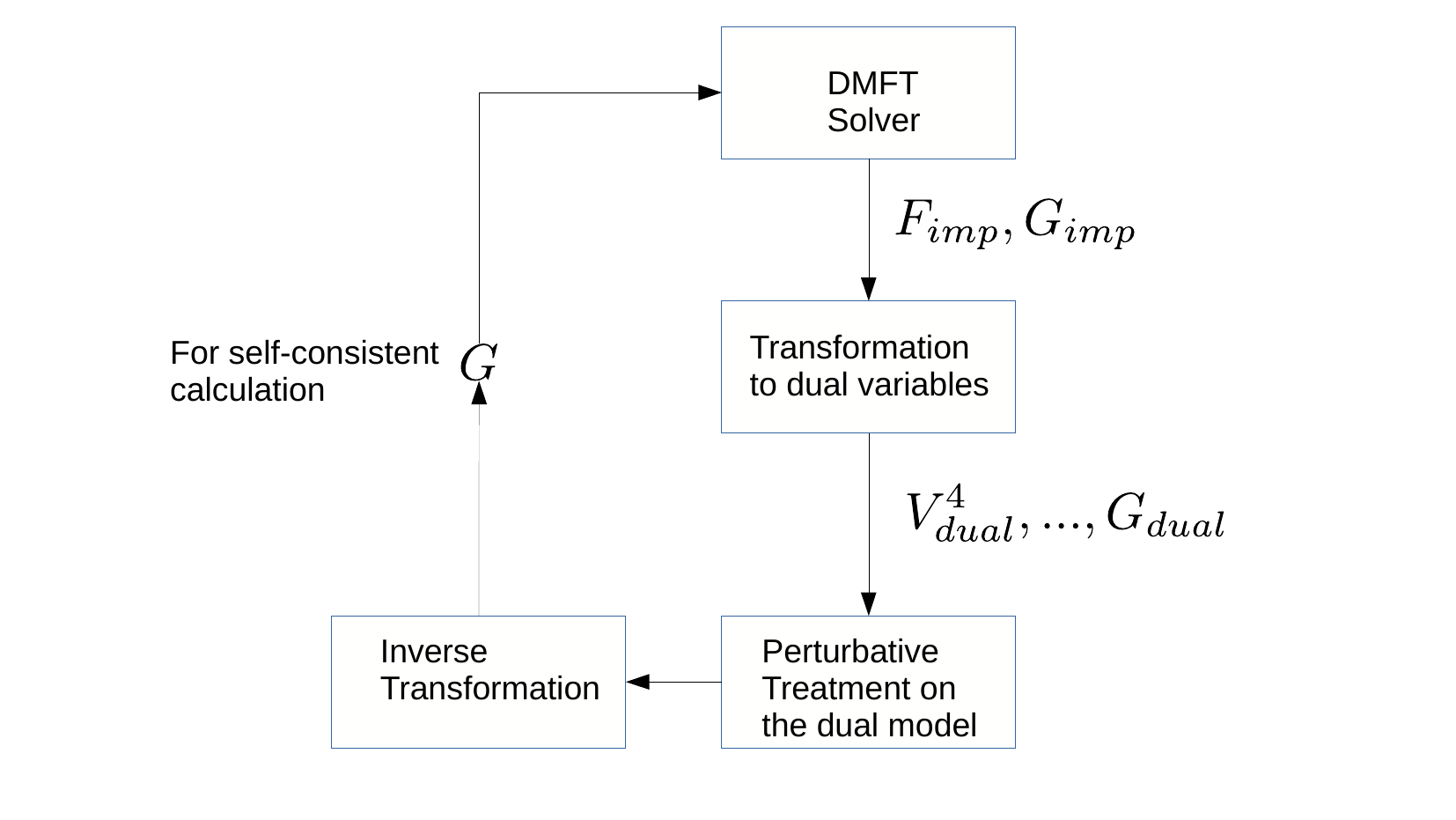}
\caption{Algorithm for the dual fermion method. {\bf 1.} The DMFT solver provides the Green function and the full (reducible) vertex function. {\bf 2.} The Green function and the irreducible vertex function are used to construct the dual fermion bare Green function and the dual potential. {\bf 3.} A perturbative method or the parquet method can be used to calculate the self-energy and the Green function of the dual fermions. {\bf 4.} Inverse transform of the Green function from dual fermions to physical fermionic degree of freedom. The results can be read out after this step. For self-consistent calculation. {\bf 5.} The Green function can be coarse-grained to obtain the bath Green function of the DMFT impurity problem. 
} 
\label{fig:df_algorithm}
\end{figure}

An explicit expression for the dual potential is obtained by expanding both sides of this equation and comparing the resulting expressions order by order.
The dual potential to lowest order reads
\begin{eqnarray}
V[f^{*},f] &&=\frac{1}{4}\sum_{\omega\omega'\nu}
\sum_{\sigma_1,\sigma_2,\sigma_3,\sigma_4}
\gamma_{\sigma_1,\sigma_2,\sigma_3,\sigma_4}(i\omega,i\omega',i\nu) \nonumber\\
\nonumber\\
&&\times f_{\omega+\nu,\sigma_1}^{*} f_{\omega,\sigma_2}
f_{\omega',\sigma_3}^{*}f_{\omega'+\nu,\sigma_4} +\cdots.
\label{eq:V_dual}
\end{eqnarray}
Where $\gamma$ is the DMFT reducible or full vertex.
Thus, nonlocal correlations are addressed by solving the many-body problem with bare Green function $g$ and interaction potential $V$. 
The lattice fermions can finally be integrated out to produce an action that only depends on the dual fermions:
\begin{equation}
 S_D[f^*,f] = -\sum_{\omega, k, \sigma} f^*_{\omega, k, \sigma} G^{d,0}_{\sigma}(i\omega, k)f_{\omega, k, \sigma} + \sum_{i} V[f^*_i, f_i].
 \label{eqn:action_ff}
\end{equation}
With the bare dual fermion Green function defined by:
\begin{equation}
 G^{d,0}_{\sigma}(i\omega, k) = -\frac{g^2(i\omega)}{[g(i\omega) + \left(\Delta(i\omega) -h_k\right)^{-1}]}.
\end{equation}
The action of Eq.~(\ref{eqn:action_ff}) is the tool to account for nonlocal correlations. It can be treated using diagrammatic perturbation theory. In this context, the interaction potential is usually truncated to the 4-point vertex.
For most practical calculations, higher order terms of the dual potential are truncated, \cite{Gukelberger_etal_2017} though they may have non-negligible effect. \cite{Ribic_etal_2017,Katanin_2013}
However, the formalism is shown, by construction, to be convergent both in the strong coupling and in the weak coupling regime. The process for solving the formalism follows a typical diagrammatic procedure. The impurity Green function and the vertex are obtained from the DMFT calculation. The impurity Green function is used to evaluate the Dual fermions non-interacting Green function. The vertex and the non-interacting Green's function are then used for a self-consistent diagrammatic solution with a given subset of all the allowed diagrams. This solution produces the dressed dual fermion Green's function. The dual Green function can subsequently be used to evaluate the lattice Green function. Alternatively, phase transitions can be studied directly using the dual fermion diagrams since the instability identified in the dual fermions space is found to be equivalent to that of the lattice Green function. In general, upon solving the dual fermion problem, a new expression for the impurity self-energy or the impurity hybridization can be extracted and fed back into the impurity solver and the entire procedure repeated iteratively until convergence. The algorithm of the dual fermion method is summarized in Fig.~\ref{fig:df_algorithm}.

The dual fermion formalism, as discussed above, was initially introduced to add nonlocal correlations to the DMFT result. It was subsequently extended to the DCA.\cite{S_Yang_2011_DFDCA, S_Iskakov_2016} In this context, it serves as a way to address intermediate length scales beyond the short ones that are treated by the cluster solver of the DCA formalism.

Because the dual fermions concept is rather general, it has been the subject of numerous developments. \cite{Hafermann_etal_2009a,Hafermann_etal_2009b,Iskakov_etal_2016,Krivenko_etal_2010,Astretsov_etal_2020,Hafermann_etal_2008} A notable generalization is the treatment of disorder with this formalism. \cite{h_terletska_13,Haase_etal_2017} The effect of disorder in correlated systems has long been an important outstanding problem in condensed matter physics, particularly for the two dimensional case. While experimentally available systems such as semi-conductors usually involve long range Coulomb interactions, the problem of a perhaps simpler Anderson-Hubbard model which have both interaction and disorder at local sites still represents an outstanding challenge which has attracted a lot of attention. 

Another important development is the treatment of nonlocal interactions, such as that of the extended Hubbard model with nearest neighbor interaction. The inclusion of nonlocal interaction opens up the possibility for interesting physics such as charge density wave and the more exotic bond order wave. The dual fermions method has also been generalized for the extended dynamical mean field theory (EDMFT) to the dual bosons theory as we discuss in the following section. \cite{Rubtsov_etal_2012}

The dual fermion method has additionally been used to generalize the real space DMFT to the real space dual fermion method which allows the study of systems with open boundary condition. \cite{Takemori_etal_2016}

\section{Dual Bosons, extension of EDMFT}
\label{sec:DualBosons}

One can apply ideas similar to those of the dual fermions for DMFT to EDMFT. \cite{Rubtsov_etal_2012} As discussed in section \ref{sec:EDMFT}, the nonlocal part of the EDMFT is not limited to the hopping term, but also includes interaction terms. \cite{Si_Smith_1996} For this problem, the action can be written as: 

 \begin{eqnarray}
S\left[ c^*, c \right] = \sum_{\omega, k, \sigma}c^*_{\omega,k,\sigma}\left[i\omega + \mu -h_k \right]c_{\omega,k,\sigma}  \nonumber \\
+\frac{1}{2}\sum_{\omega,k} V(k)n_{\omega,-k} n_{\omega,k} +
\sum_i S_{local}[c^*, c].
\label{eqn:DB_action}
\end{eqnarray}

The next step is to introduce a Hubbard-Stratonovich transformation as in the dual fermion formulation for the nonlocal bilinear term of the kinetic energy. In addition, the nonlocal density-density potential energy term is decoupled by another Hubbard-Stratonovich transformation for the bosonic charge density. In parallel with the dual fermion method, the original fermionic degrees of freedom, and the bosonic charge density can formally be integrated out exactly, yielding a new effective potential given by the vertices of the impurity model. The action becomes

\begin{eqnarray}
&&S_D[f^*,f] = -\sum_{\omega, k, \sigma} f^*_{\omega, k, \sigma} G^{d,0}_{\sigma}(i\omega, k)f_{\omega, k, \sigma}  \nonumber \\
&&-\sum_{\omega, k} \phi^*_{-\omega, -k} D^{d,0}(i\omega, k)\phi_{\omega, k} +
 \sum_{i} V[f^*_i, f_i,\phi_i].
\end{eqnarray}

Similar to the dual fermions, the potential of the dual variables can be expanded and truncated at finite order of the vertex functions of the impurity problem. Perturbative methods can be used to solve the  effective problem expressed in terms of the dual variables. The method has been applied to problems with nonlocal interaction, such as the extended Hubbard model and models with long range Coulomb coupling. \cite{vanLoon_etal_2014,Stepanov_etal_2016,vanLoon_etal_2016,Vandelli_etal_2020,Peters_etal_2019,vanLoon_etal_2014}

\begin{figure}[htbp]
\includegraphics*[width=10.0cm, height=6.0cm]{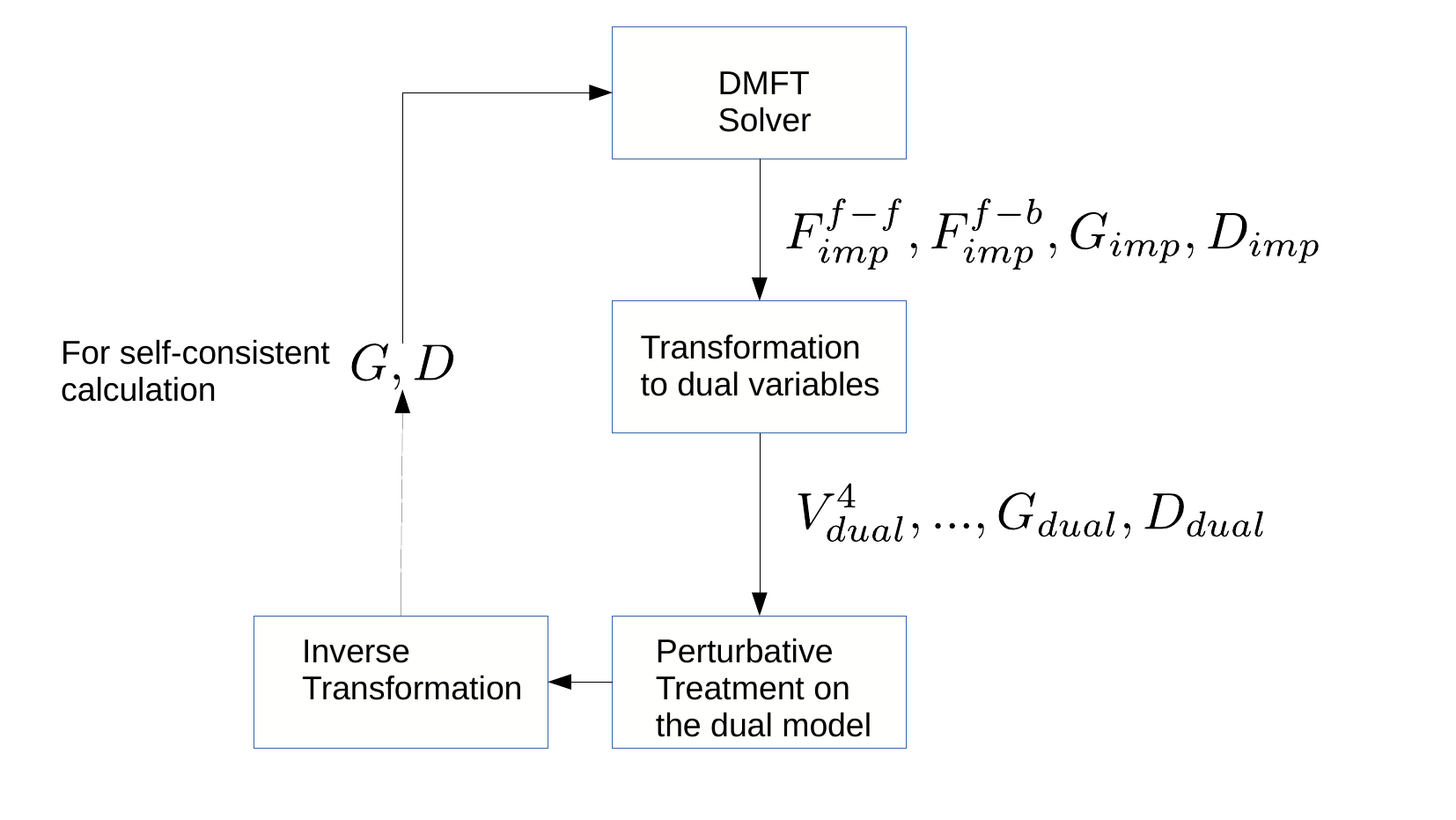}
\caption{Algorithm for the dual boson method. This algorithm is parallel to the dual fermion method, except that both the bosonic Green function and the dual potential with the bosonic degrees of freedom are needed. 
{\bf 1.} The DMFT solver provides the Green function and the reducible vertex functions. {\bf 2.} The Green function and the reducible vertex functions for both fermion-fermion vertex ($F^{f-f}_{imp}$) and fermion-boson vertex ($F^{f-b}_{imp}$) from the DMFT solver are used to construct the Green functions ($G_{dual}$ and $D_{dual}$) and the dual potential ($V_{dual}$). {\bf 3.} Perturbative methods can be used to calculate the self-energy and the Green function of the dual fermions and dual bosons. {\bf 4.} Inverse transformation of the Green function from dual fermions and dual bosons to physical fermionic and bosonic degrees of freedom. The results can be read out after this step. For the full self-consistent calculation, {\bf 5.} The Green function can be coarse-grained to obtain the bath Green function of the DMFT impurity problem.
}
\label{fig:db_algorithm}
\end{figure}

The algorithm of the dual bosons method is summarized in Fig.~\ref{fig:db_algorithm}.

\section{GW approximation with three-particle irreducible vertex}
\label{sec:GW}

A conventional scheme to address quantum fluctuations beyond Hartree-Fock is the so-called GW approximation. It was developed for solutions of the electron gas problem in the 50's. Due to its simplicity, it has been widely adapted in density functional theory calculations and it can be derived from many-body perturbation theory. The form of the self-energy in the GW approximation is kept as that of the Hartree-Fock approximation, but the interaction, originally just the Coulomb term, is dynamically screened. \cite{Quinn_Ferrell_1958,DuBois_1959a,DuBois_1959b,Hedin_1965,gw}

Recent proposals have utilized the properties of the GW approximation and extended it by introducing a dynamical three point vertex function. The method can be derived from the partial bosonization of the electron-electron interaction via the Hubbard-Stratonovich transformation in different channels. The effective action becomes that of an electron-boson coupling problem. The bosonic fields from the decoupling of the electron-electron interaction can be considered as the electrons coupled to the charge and spin fluctuations. This is equivalent to solving the self-energy in the Hedin's equation with the vertex correction \cite{Hedin_1965}. Instead of solving the electron-boson vertex with spatial dependence, the vertex is calculated via an effective impurity problem similar to that of DMFT. The spatial dependence of both the fermion and the boson self-energies is generated by the fermion ($G$) and the boson ($D$) Green functions. 
This generalization was introduced by Aryal and Parcollet as the Triply Irreducible Local Expansion (TRILEX). \cite{Ayral_Parcollet_2015,Vucelse_etal_2017} The method can be formally derived from a functional of the vertex given by three-particle irreducible diagrams \cite{c_dedominicis_64,DeDominicis_Martin_1964}.

\begin{figure}[htbp]
\includegraphics*[width=10.0cm, height=6.0cm]{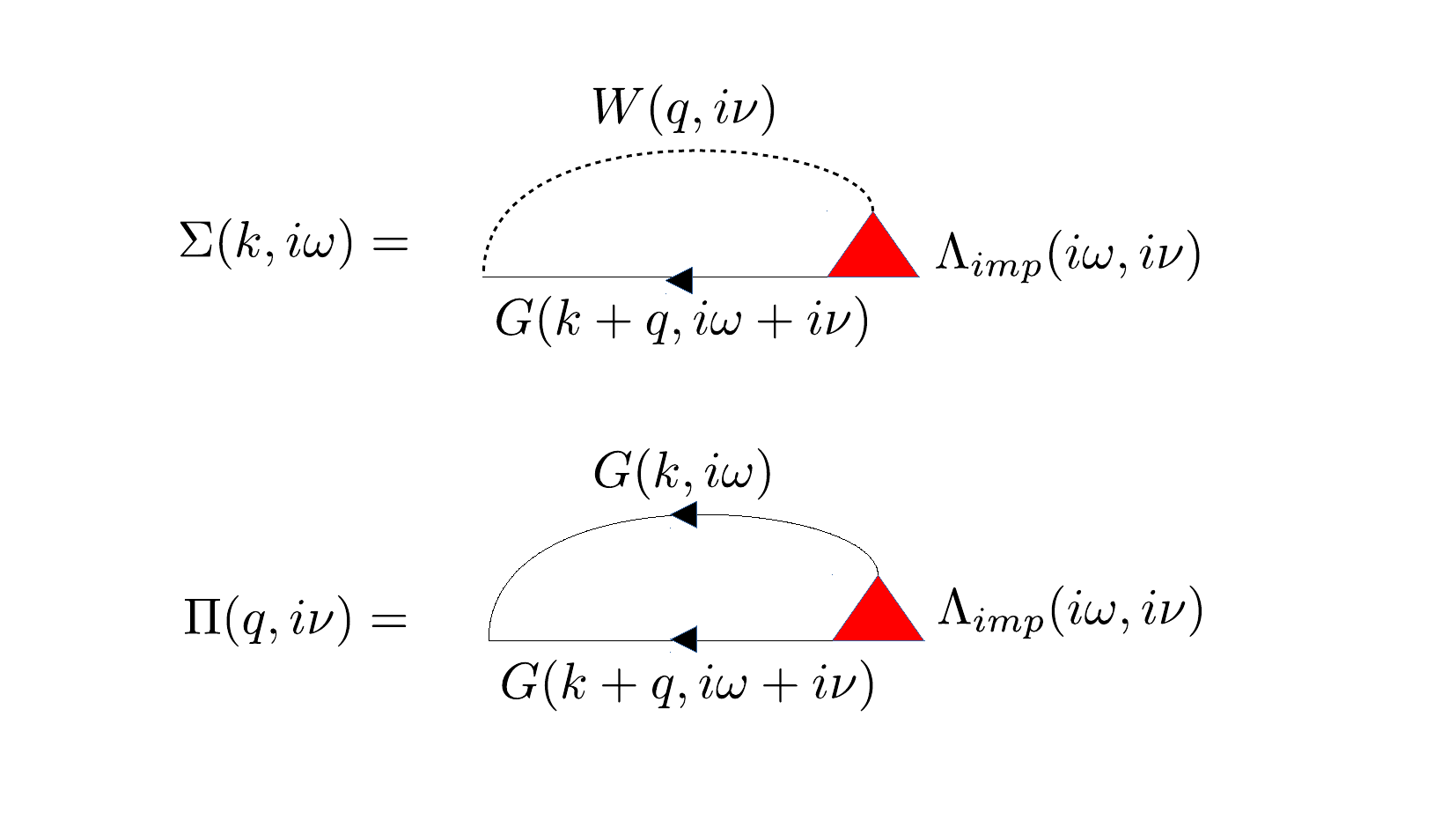} 
\caption{Diagrams for generating momentum dependence for the self-energy of the single particle function and the polarization of the two particle function.
} 
\label{fig:trilex_diagrams}
\end{figure}

The key step for incorporating nonlocal correlations is to consider the polarization and the electron self-energy with the local boson-fermion vertex from the numerical solution of the impurity problem, see Fig.~\ref{fig:trilex_diagrams} for the diagrams. The rest of the algorithm is parallel to that of the EDMFT. The updated self-energy and the polarizability are coarse-grained and fed back into the EDMFT effective impurity problem. 

\begin{figure}[htbp]
\includegraphics*[width=10.0cm, height=6.0cm]{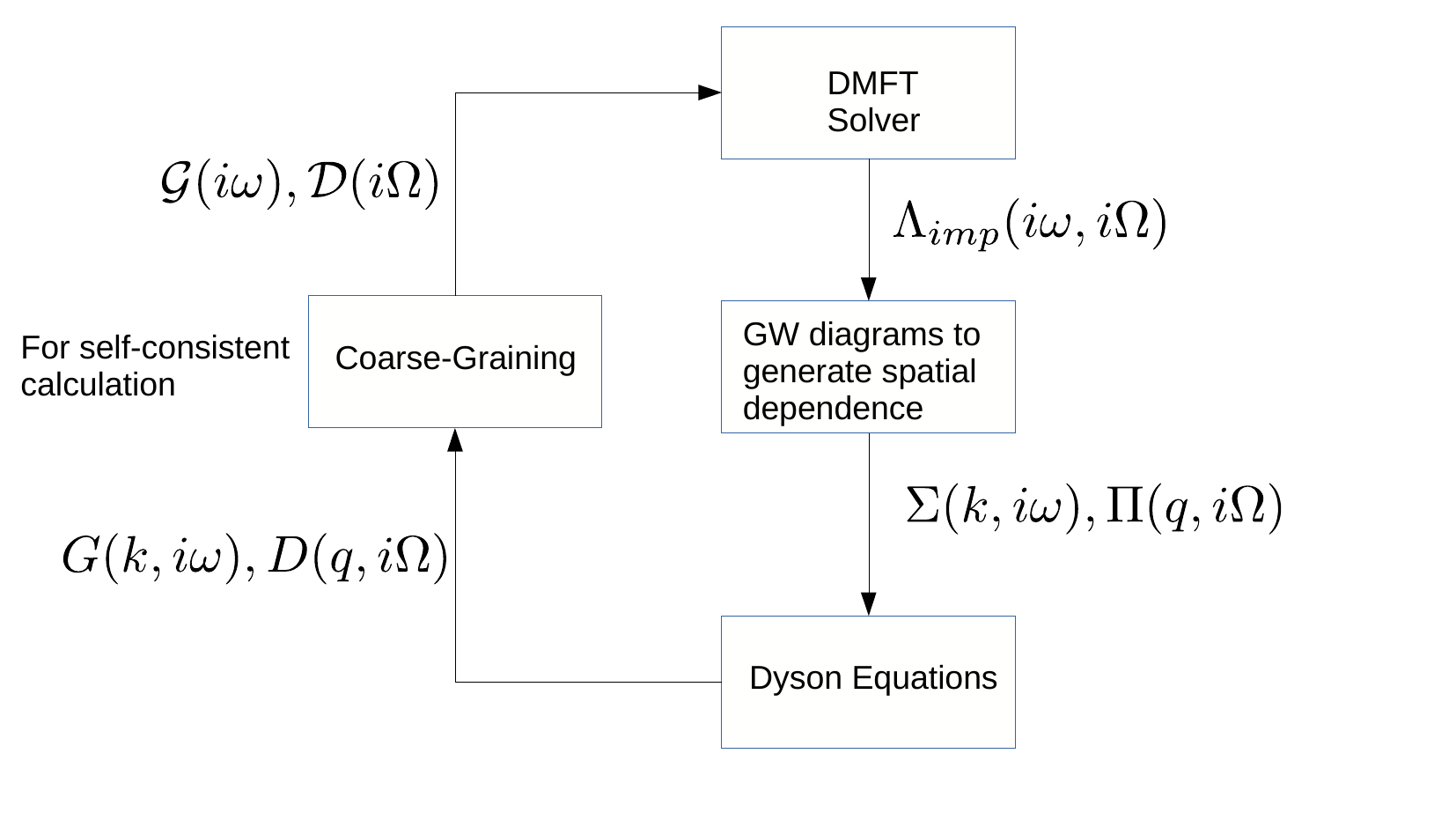}
\caption{Algorithm for the TRILEX method. {\bf1.} The DMFT solver provides the boson-fermion  vertex. {\bf 2.} The Hartree diagrams of the GW approximation are calculated with the vertex from the DMFT for both the self-energy and the polarization to generate spatial dependence. {\bf 3.} The Dyson equation is employed to calculate the Green function for both the fermions ($G$) and the bosons ($D$). {\bf 4.} The coarse-grained Green functions can be fed back to the effective DMFT impurity problem. Repeat from step 1 until convergence is attained.} 
\label{fig:trilez_algorithm}
\end{figure}

The algorithm for the TRILEX method is summarized in Fig.~\ref{fig:trilez_algorithm}.

\section{Functional Renormalization Group}
\label{sec:FRG}

The functional renormalization group for fermionic systems evolved from approaches used 
in the one dimensional Luttinger liquid.
 \cite{Voit_1995,Tomonaga_1950,Luttinger_1963,Giamarchi_2003} The full vertex in this context is usually denoted as ``g", for this reason, the method is often referred as g-ology. \cite{ Menyhard_Solyom_1973,Solyom_1973} From the renormalization group perspective, the frequency dependence of the vertex function is not the most relevant contribution. Thus for systems that do not have explicit frequency dependence in the bare vertex, such as the electron-phonon coupling, the frequency or time dependence is often neglected. \cite{Shankar_1994} 

The generalization of g-ology beyond one dimension gained increased attention after the work by Shankar. \cite{Shankar_1991,Shankar_1994} Traditional implementations of Wilsonian renormalization group only consider the flow of a handful of coupling constants, which is physically the case for one dimensional systems as there are only two Fermi points instead of a surface. The functional formalism for all coupling constants is considered when the method is generalized to high spatial dimensions.

The idea of performing renormalization group on a system with an extended Fermi surface was suggested by Anderson. \cite{Anderson_book_1984} Benfatto and Gallavotti \cite{Benfatto_Gallavotti_1990}, and Feldman and Trubowitz \cite{Feldman_Trubowitz_1990,Feldman_Trubowitz_1991} studied the stability of the Fermi liquid against perturbations. Effective theories based on the idea of renormalization group for Fermi liquid were derived by Polchinski, \cite{Polchinski_1992} and for superconductors by Weinberg. \cite{Weinberg_1993} 

Most studies in the late 90's and early 00's focused on the two dimensional Hubbard model. \cite{Zanchi_Schulz_2000,Zanchi_Schulz_1998,Honerkamp_etal_2001,Halboth_Metzner_2000a,Halboth_Metzner_2000b,Kampf_Katanin_2003,Tsai_Marston_2001} Two important developments paved the way for using FRG to improve DMFT solutions. First, originally, the conventional renormalization group uses an energy cutoff. \cite{Shankar_1994} This is clearly not a unique choice, 
and other cutoffs such as temperature, interaction, and even an hybridization cutoff for the impurity problem have been proposed and implemented. \cite{Honerkamp_etal_2004,Honerkamp_Salmhofer_2001,Kinza_Honerkamp_2013}  The second 
development is the implementation of the frequency dependent vertex. This is largely motivated by the interest in studying more complicated  electron-phonon coupling models. \cite{Honerkamp_Salmhofer_2005,
Tsai_etal_2005} When only electron-phonon coupling is considered, 
the frequency dependence, even at the single particle level, is not considered. That is, the self-energy is not renormalized. The situation changes when explicitly retarded interactions are  considered in models with electron-phonon coupling or local disorder, such as the Anderson impurity model. \cite{Tam_etal_2007a,Tam_etal_2007b,Sedeki_etal_2000,Karrasch_etal_2008}

It was realized early in the study of the frequency dependent vertex that the method can be used to study mean field fluctuations. Attempts have been made to use it for the study of Gaussian fluctuations of the slave boson mean field solution for the t-J model. \cite{Kotliar_Liu_1988} 
Progress in this research direction has been limited because the conventional assumption that the bandwidth should be larger than the interaction is not met for a wide range of parameters of the slave boson t-J model. 

Related ideas have been revived in the past few years to consider mean field fluctuations. Instead of a static mean field solution, the solution of the DMFT is considered. \cite{Tranto_etal_2014} The technique developed as a result can be directly applied to build up the spatial dependence from the DMFT or DCA solution. The main idea is that the bare vertex is replaced by the full vertex from the DMFT or DCA, and the self-energy is set to that of the DMFT or DCA for the initial conditions of the renormalization flows.

From a general perspective, the renormalization group can be viewed as summing up diagrams that are often referred to as forming the leading divergence. \cite{Yakovenko_1996} In particular if second order perturbation is used in the renormalization group calculation, the diagrams generated have the same topology as those of the parquet method discussed above. 

There is however a subtlety. The integration over the internal frequency and momentum are not the same as those of the parquet method. The difference depends on the different cutoff schemes being used, but the summation is never performed over the full range of frequency and momentum, instead it is done on a shell of the energy range, and iteratively approaches the desired low energy or low temperature manifold. 

It has long been believed that the leading divergences of the one-loop FRG and the parquet should be equivalent. \cite{Diekmann_Severin_2020,Solyom_1973,Zheleznyak_etal_1997} A recent proposal of a multi-loop flow equation for the four-point vertex framework showed that the FRG flow consisting of successive one-loop calculations is equivalent to a solution of the parquet equations. This further supports the idea of FRG as a possible alternative to solving the parquet equations.\cite{Kugler_vonDelft_2018a,Kugler_vonDelft_2018b}

Given the same topology of the diagrams, one might expect that the leading divergence among these two methods should be the same even outside of the multi-loop setting. Therefore, for the purpose of looking for the instability from metallic to ordered phase, these two approaches should be expected to give the same result. On the other hand, for the metallic phase with no proximity to an instability, the results are not naively equivalent. 

The clear advantage of the functional renormalization group is the ease of the numerical calculations. Unlike, the parquet or even the simplified dynamical vertex approximation, the functional renormalization group works directly on the full reducible vertex. The numerical solution does not involve solving self-consistent equations. It is given by the flow of the full vertex as the cutoff is lowered to the desired energy or temperature. The formulation is represented in term of ordinary differential equations. 

The main challenge of the parquet method is the instability of the numerical solution. Even if we assume the existence of a unique solution, a robust method to attain this solution remains highly non-trivial. This is ultimately the main issue with methods that seek self-consistent solutions. Unlike the dynamical mean field theory in which only single particle quantities are involved, the two-particle methods, full parquet or simplified forms, involve solving for the vertex through self-consistent equations. Divergence in the vertex function is expected to occur as the temperature is lowered. Therefore, equations with variables spanning a large range of values over many orders of magnitude have to be solved. This is clearly a non-trivial task from the perspective of numerical simulations. Although one can solve the equations at high temperature, it is not always clear how far the solution can be pushed down in temperature. Contrary to this, the functional renormalization group method sums up the same set of diagrams without the necessity of solving relevant equations self-consistently. The divergence is approached step by step rather than via shooting as is done in the self-consistent solution. 

While the advantage of functional renormalization group from the point of view of numerical stability is clear, the justification of its usage for nonlocal corrections with the vertex function from the DMFT solution is not obvious. The conventional wisdom of renormalization group calculations is that the bandwidth should be large and the interaction is a small parameter. The full vertex from DMFT is not necessarily small compared to the effective bandwidth. Therefore, the conventional wisdom of justifying the low order expansion is not incontrovertibly fulfilled. Moreover, these couplings generate self-energy corrections, which have been shown to be important for studying systems with retardation effects, leading to the renormalization of the Fermi velocity and quasiparticle lifetime. Thus, these couplings have to be kept even though they are often ignored in the non-retarded systems. In brief, the effective system being solved is retarded although the original Hubbard model is not.

To leading second order expansion, the renormalization group equations for the scale $(\Lambda)$ dependent full vertex function, $F_{\Lambda}(k_1,k_2,k_3)$, \cite{Zanchi_Schulz_1998,Zanchi_Schulz_2000} and the self-energy, $\Sigma_{\Lambda}(k)$ for a spin rotational invariant two-body interacting system, are given by: 
\begin{widetext}
\begin{eqnarray}
\partial_{\Lambda} F_{\Lambda}(k_1,k_2,k_3) =
&-& \int dp \partial_{\Lambda}
[G_{\Lambda}(p)G_{\Lambda}(k)] F_{\Lambda}(k_1,k_2,k)
F_{\Lambda}(p,k,k_3)
\nonumber \\
&-&\int dp \partial_{\Lambda}
 [G_{\Lambda}(p)G_{\Lambda}(q_1)]
F_{\Lambda}(p,k_2,q_1) g_{\Lambda}(k_1,q_1,k_3)
\nonumber\\
&-& \int\!dp \partial_{\Lambda}
 [G_{\Lambda}(p)G_{\Lambda}(q_2)][-2g_{\Lambda}(k_1,p,
 q_2)F_{\Lambda}(q_2,k_2,k_3)
\nonumber\\
&+&F_{\Lambda}(p,k_1,q_2) F_{\Lambda}(q_2,k_2,k_3)+F_{\Lambda}(k_1,p,q_2)
F_{\Lambda}(k_2,q_2,k_3)],
\end{eqnarray}
\begin{eqnarray}
\partial_{\Lambda} \Sigma_{\Lambda}(k) \!=\! 
-\!\int\!dp \partial_{\Lambda} [G_{\Lambda}(p)]
[2F_{\Lambda}(p,k,k)\!-\!F_{\Lambda}(k,p,k)],
\end{eqnarray}
\end{widetext}
where $k=k_1+k_2-p$, $q_1=p+k_3-k_1$, $q_2=p+k_3-k_2$,
$\int dp=\int
d\mathbf{p}\sum_{\omega}1/(2\pi\beta)$, and $G_{\Lambda}$ is the self-energy corrected propagator
at cutoff $\Lambda$.

\begin{figure}[htbp]
\includegraphics*[width=9.0cm, height=5.0cm]{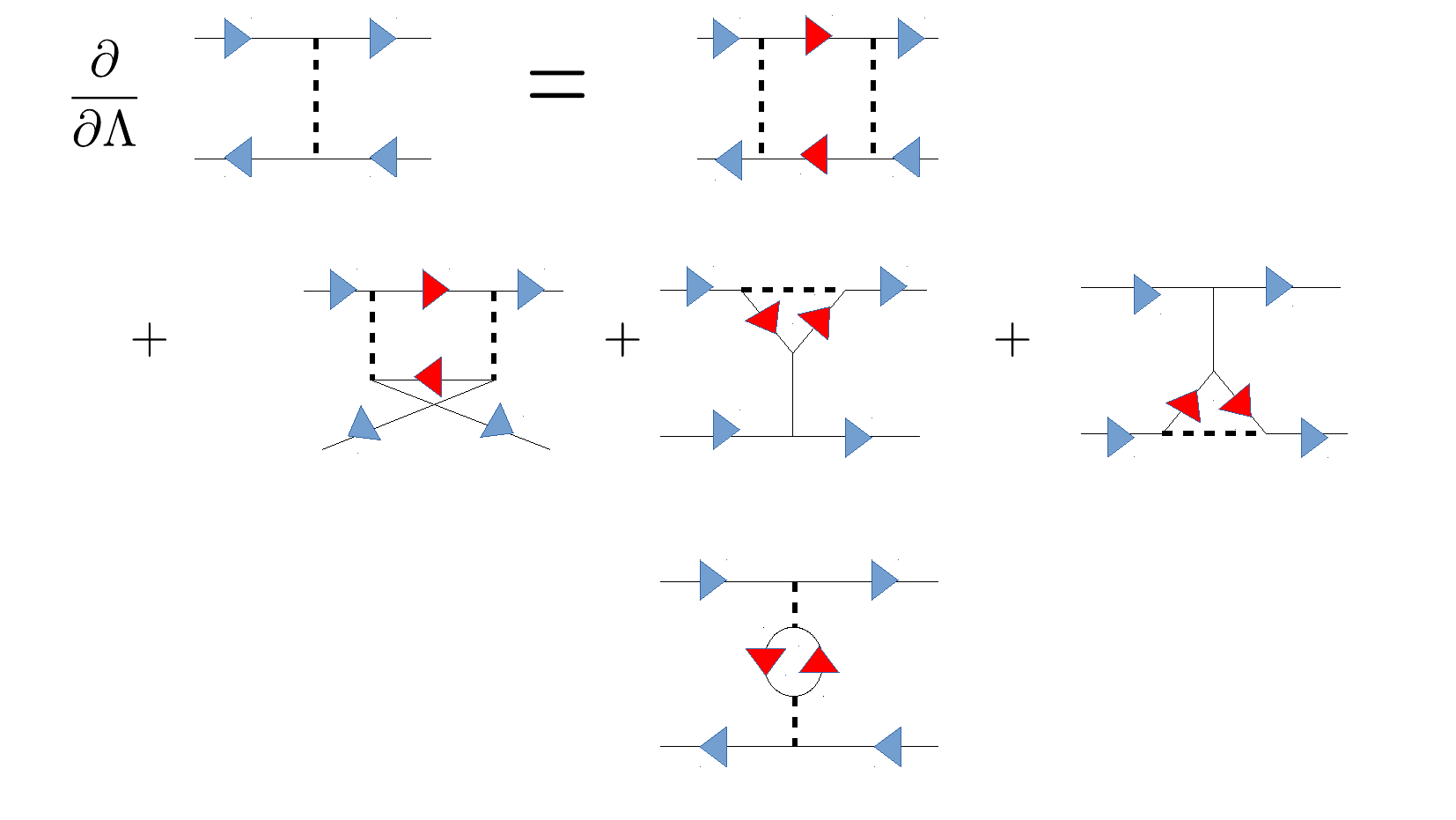} 
\caption{RG equations for the full vertex. There is a pair of red arrows which represent the derivative of the product of propagators with respect to the RG scale $\Lambda$.} 
\label{fig:frg_diagrams}
\end{figure}

The RG equation can be presented in terms of diagrams. For a non-retarded system, the low energy instability can be obtained from the renormalization flow of the couplings, and different phases can be identified by the fixed points corresponding to the relevant spin and charge modes. 
One can explicitly construct the flows of the susceptibilities of different order parameters. For example the pairing susceptibility, $\chi^{\delta}(\mathbf{k},\omega)$, is defined by: 
\begin{eqnarray}
\label{SC_susceptibility}
\chi^{\delta}_{\Lambda}(0,0)\!\!=\!\!\int\!\!
\int d p_1 d p_2
\langle c_{p_{1},\downarrow}c_{-p_{1},\uparrow}c_{-p_{2},\uparrow}^{\dagger}c_{p_{2},\downarrow}^{\dagger}\rangle_{\Lambda}.
\end{eqnarray}
The RG equations are: 
\begin{eqnarray} \label{SC_susceptibility RGE}
\partial_{\Lambda} \chi^{\delta}_{\Lambda}(0,0) = \int d p \partial_{\Lambda}[
G_{\Lambda}(p)G_{\Lambda}(-p)](Z^{\delta}_{\Lambda}(p))^2,
\end{eqnarray}
\begin{eqnarray}
\partial_{\Lambda} Z^{\delta}_{\Lambda}(p) = -\int dp^{\prime}\partial_{\Lambda} [
G_{\Lambda} (p^{\prime})G_{\Lambda}(-p^{\prime}) ]\\ \nonumber Z^{\delta}_{\Lambda}(p^{\prime})g_{\Lambda}(p^{\prime},-p^{\prime},-p,p).
\end{eqnarray}

The function $Z_{\Lambda}^{\delta}(p)$ is the effective vertex in the definition for the susceptibility $\chi_{\Lambda}^{\delta}$.
The RG equations for susceptibilities are solved with initial condition $\chi^{\delta}_{\Lambda=\Lambda_{0}}(0,0)=0$. The dominant instability in the ground state is given by the most divergent susceptibility by solving the renormalization equations numerically. Similar equations can be derived for other susceptibilities.  

For the FRG boosted DMFT approach, DMF$^2$RG, \cite{Tranto_etal_2014}
the initial condition for the full vertex functions and the self-energy are both given by the DMFT. \cite{Tranto_etal_2014} The scale dependent bare propagator is defined as an interpolation between the DMFT propagator and the bare lattice propagator as
\begin{eqnarray}
G_{\Lambda}^{0}(\mathbf{k},i\omega)^{-1} =\Lambda G_{imp}^{0}(i\omega)^{-1}  \\ \nonumber + (1-\Lambda) G_{lattice}^{0}(\mathbf{k},i\omega)^{-1}.
\end{eqnarray}

Since one can treat the functional renormalization group as a vehicle for summing diagrams, it can in principle be applied on any effective fermionic interacting system. For example, it has recently been used as a solver for the dual boson method. \cite{Katanin_2019} FRG on auxiliary fermion or dual fermion has also been considered. \cite{Wentzell_etal_2015,Katanin_2015}

\begin{figure}[htbp]
\includegraphics*[width=10.0cm, height=6.0cm]{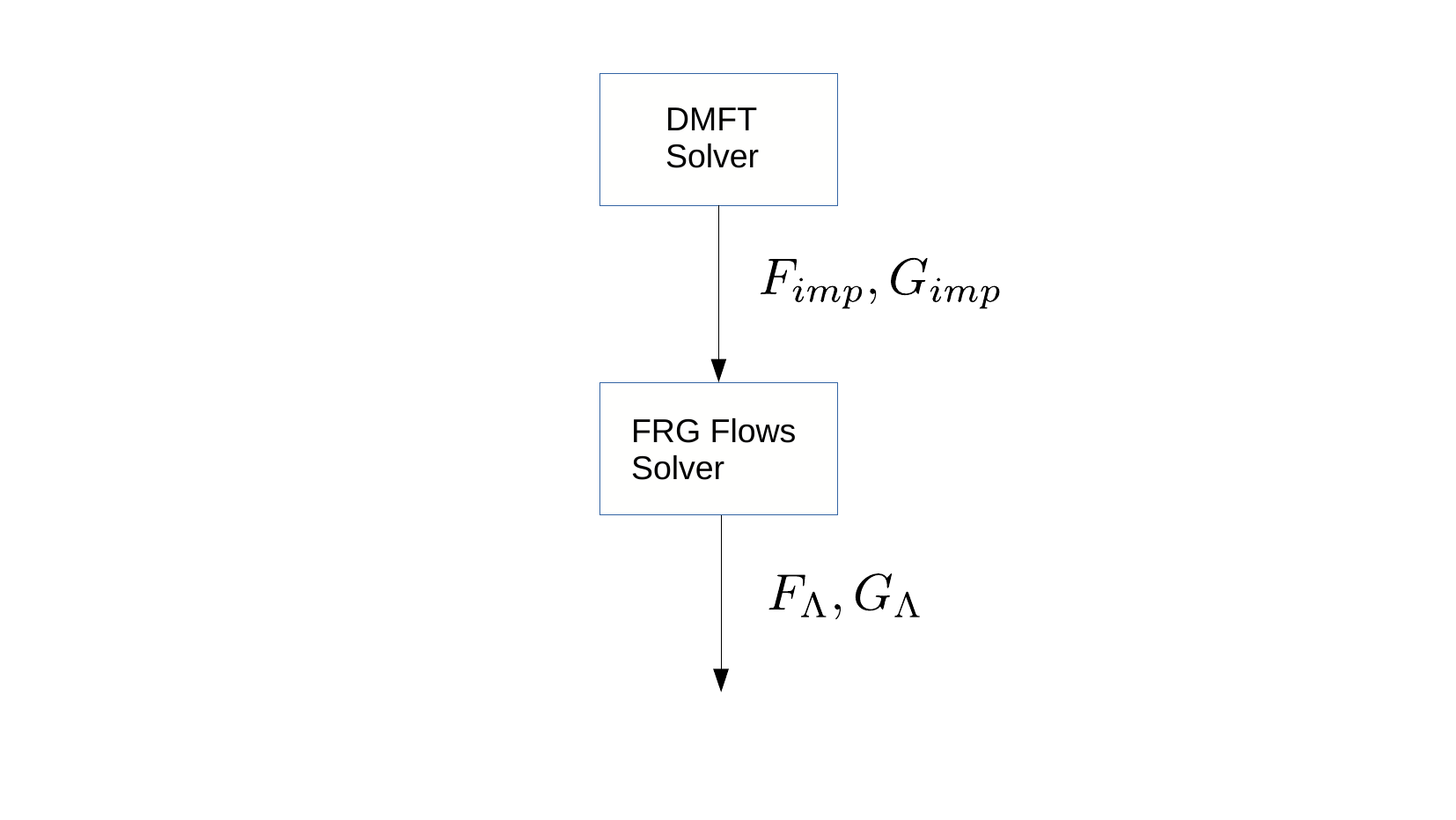}
\caption{Algorithm for the
functional renormalization group boosted DMFT. {\bf 1.} The full reducible vertex and the Green function are obtained from the DMFT solver. {\bf 2.} They are fed into the FRG method as the initial conditions. {\bf 3.} Solve the flow equations for the FRG to obtain the Green function, full vertex, and susceptibilities.
} 
\label{fig:frg_algorithm}
\end{figure}

The DMF$^2$RG algorithm is summarized in Fig. \ref{fig:frg_algorithm}. For the derivation of the FRG formulation in the context of strongly correlated systems we refer the readers to references [\onlinecite{Zanchi_Schulz_1998,Honerkamp_Salmhofer_2001,Shankar_1994,Binz_etal_2003}]. Details of the implementation and approximations of the DMF$^2$RG can be found in reference [\onlinecite{Tranto_etal_2014}].

\section{Numerical methods to represent vertex functions}
\label{sec:Numerics}

The various methods described above in principle provide nonlocal corrections to the conventional dynamical mean field theory or additional nonlocal corrections in the case of the dynamical cluster approximation or the cellular dynamical mean field theory. The ultimate goal is to attain a better approximation to the exact solution. The physically most interesting regimes are often in the intermediate coupling away from the trivial limits that could serve as a good basis for perturbative treatments. After all, all considered methods are based on some truncation with respect to the interaction or 
more complicated objects as those in the ``dual" variables approach. Moreover, the actual numerical implementation is sometimes rather challenging. Unlike DMFT where only single particle quantities are involved in the self-consistent equations, the storage requirement of two particle vertex functions is increased by two powers of the space-time grid size. Simply storing those vertices is in itself a rather difficult task. Of course, that largely depends on the interaction and the temperature range. One can naively expect that  more fine resolution in the space-time grid is needed for intermediate interactions and low temperatures.
For high temperature and very weak or very strong coupling, a rather sparse grid for discretizing the space-time functions could be sufficient. 

The above storage problem can be resolved to an extent in today's clusters with tens of thousands of computing nodes. How to efficiently manipulate an object with such large memory requirement, and involving all-to-all memory swaps, is an active research problem in computer science. \cite{Wagle_etal_2018,k_tam_13} Altogether, with the improvement of computer hardware and better implementations, the storage problem can be mitigated. 

The main issue that is ultimately common to most of these methods is the convergence to a solution. There is in general no guarantee that a self-consistent solution exists. While 
even in the single particle self-consistent method, there is no guarantee that there is a uniform convergence to the solution. The situation is more acute in theories that require two particle self-consistency. The vertex function is a measure of the instability towards an ordered phase, therefore it should become singular as the instability is approached. One can often work in the range of interactions and temperatures where singularities are far from being reached. However, this may altogether defeat the purpose of the methods as the regimes near the instabilities are usually those where corrections to the dynamical mean field theory are of most interest. The challenge of finding a stable numerical solution of the parquet formalism was already discussed in the early days when the formalism was applied to the single impurity Anderson model. \cite{Bickers_homotopy,n_bickers_92} It is fair to say that a  universally reliable approach for a solution has yet to be found.

For DMFT, besides the simple iterative method, the more powerful Broyden method which utilizes the gradient of the hybridization function in the impurity problem has been implemented and is occasionally used in cases where convergence is difficult to achieve. \cite{Ztiko_2009} For the two-particle theories, given the complexity of the equations being solved, more sophisticated methods have been attempted. One of them is the homotopy method, in which a known convergent solution is relaxed to hopefully lead to a solution for another temperature or interaction strength. Practically, the methods are not in general easy to apply and convergence is not guaranteed. 

The above difficulties in solving for the two-particle vertex function self-consistently may give an advantage to  FRG based methods in which no self-consistent solution is sought. The solution is obtained not by solving non-linear integral self-consistent equations, but rather by solving differential equations with initial conditions. This allows more flexibility in the numerical solution. 

For substantial progress to be accomplished, it is essential that numerical methods perform well while, at the same time, requiring reasonable computational resources. Although storing the full vertex is a daunting task, the amount of information it actually contains is, in practice, not very large. At least for the weak coupling case, the vertex functions contain very small entropy in the sense that it can be compressed numerically to a large extent and still retain most of the information. It has been suggested for a long time that, a possible route to storing the vertex function is to use the spectral representation. \cite{Shvaika_2016,Shvaika_2006} A clear advantage in the spectral representation is that the high frequency information is built in the representation.
The spectral representation has been further explored in recent studies. \cite{Shinaoka_etal_2018,Shinaoka_etal_2017} 

Various efforts have also been devoted to understanding the frequency structure of the vertex function. These may help with new ideas on approximation schemes for the vertex function. \cite{j_kunes_11,Katanin_2020,Li_etal_2016,Thunstrom_etal_2018,Wentzell_etal_2020,Rohringer_etal_2012,Chalupa_etal_2018} The latest proposal is to use a tensor network representation. \cite{Shinaoka_etal_2020}

Yet another intuitive scheme is to consider an inhomogeneous grid to represent momentum-frequency space indices. Generically, the frequency or momentum dependence is described by an interpolation scheme. This can be justified specifically for the frequency indices because the low frequency information should be more important, contains most of the information and thus requires higher resolution. Moreover the high frequency contribution can be well fitted by simple functions for convenient storage. Generally, such methods which are based on interpolations, can be seen as approximating the vertex as follows:
\begin{equation}
F(\omega_{k},\omega_{k^{'}},\omega_{q}) = 
S(\tilde{F}(\tilde{\omega}_{k},\tilde{\omega}_{k^{'}},\tilde{\omega}_{q}),\omega_{k},\omega_{k^{'}},\omega_{q} ), 
\end{equation}
where $F$ is the vertex in the uniform frequency grid $\omega$ and $\tilde{F}$ is the actual data stored in the grid of some arbitrary basis in $\tilde{\omega}$. $S$ is the interpolating function or the basis function that maps $\tilde{F}$ to $F$. 

Note that there is no inverse for such an interpolation or basis expansion. Similar ideas can also be extended to the space grid. This has been studied recently in the context of FRG and also the parquet method. \cite{Eckhardt_etal_2020,Schober_etal_2018,Honerkamp_2018,Lichtenstein_etal_2017,Eckhardt_etal_2018}  An extreme case of only retaining the transfer frequency and momentum has been proposed as a simplification of the parquet equations for analytical solutions for single impurity problems. \cite{Augustinsky_Janis_2011,Janis_Augustinsky_2007,Janis_Augustinsky_2008}

The lack of an inverse transformation to the original vertex means that the self-consistent equations are, in principle, altered by the transformation. The representation is however a controlled approximation in the sense that the larger number of basis functions or the larger number of grid points can reduce the acquired error. On the other hand, strictly speaking, the crossing symmetries of the vertex, that are a manifestation of the Pauli exclusion principle, are also broken by such interpolation or basis expansion schemes. These symmetries are one of the key features of the parquet algorithm.



A scheme that has been studied in the context of FRG is to factorize the vertex function approximately. \cite{Tranto_etal_2014} In the representation of the vertex function with two fermionic frequencies and one bosonic transfer frequency, the frequency transfer has the most dominant contribution. 
Thus, one can argue physically that the frequency or momentum transfer part of the vertex function can be factorized. This drastically reduces the computational effort and, in particular, the storage requirement. A less drastic approximation is the so-called two-level approximation, \cite{Li_etal_2016} for which the low frequency part of the vertex is calculated exactly while only the frequency transfer dependence is kept for the high frequency part of the vertex.
Another recent proposal is to consider the momentum dependence by using an expansion in terms of, say, spherical harmonics. This can help reduce the storage for the spatial dimension and thus allow studies of larger cluster sizes. \cite{Eckhardt_etal_2020} 

It is important to point out out that the corrections on top of the DMFT solution, while they are indeed quite meaningful, are often not large for most momentum-frequency points, particularly for weak or very strong interactions. It is, in this sense, appropriate to inquire what are the effects of numerical errors on a method that would, in principle, be capable of significant corrections to DMFT. Might the approximate numerical solutions overwhelm the expected solution? 
Of course, one could expect that corrections in the intermediate regime are larger, but the convergence problem in this regime may prevent a satisfactory solution. 

Overall, a number of challenges still remain for general numerical solutions. 
Algorithmic breakthroughs, advances in the representation of the vertex function and in the solvers, are needed for general converging solutions over a wider range of parameters.

\section{Exact Constraints} 
\label{sec:Constraints}

\subsection{Conserving approximation}

A milestone in the theory of condensed matter physics is the formulation of the so-called conserving approximations by Baym and Kadanoff. \cite{g_baym_61,g_baym_62} The idea of constructing approximations based on a functional of the Green function was pioneered by Luttinger and Ward, $\Phi(G)$. \cite{Luttinger_Ward_1960} It was shown that the functional derivative with respect to the Green function can be used to define the self-energy $\delta \Sigma = tr \Sigma \delta G$. Putting it together with the Dyson equation $G^{-1}=G_{0}^{-1} -\Sigma$, the correlation functions generated by the functional obey conservation laws and thus the Ward identities. 

Conventional methods that are also conserving approximations include self-consistent solutions of the Hatree-Fock, T-matrix approximation, and GW approximations. Another well known method that is also a conserving approximation is the fluctuation exchange (FLEX) approximation. 

Dynamical mean field theory and its cluster extensions, CDMFT and DCA, are both conserving approximations although the Ward identity on the lattice is strictly speaking not satisfied due to the coarse-graining procedure. Though ``conserving approximation" does not necessary imply ``better approximation", conserving approximations may be more important for  broken symmetry cases, that are not routinely studied by the DMFT. 

It has been argued that both the dual fermion and the dual boson approximations give conserving approximations for ladder summation. \cite{Rubtsov_etal_2012} That should also be the case for the dynamical vertex approximation, at least for the ladder approximation.

\subsection{Mermin-Wagner-Coleman theorem}
This theorem forbids broken continuous symmetry for a system in dimensions two or less at a finite temperature. \cite{Mermin_Wagner_1966,Hohenberg_1967,Coleman_1973} A particularly interesting case is that of the spin rotation symmetry in the two dimensional Hubbard model that should not be broken and thus no true phase transition should occur.

We are not aware of a proof that any of the methods presented in this review satisfies this constraint. There is plenty of evidence that the DMFT and its cluster extensions do show finite transition temperatures through measurements of the susceptibilities. 

A recent study by the parquet method seems to provide numerical evidence that Mermin-Wagner theorem could be satisfied. \cite{Eckhardt_etal_2020} The dual boson theory has also been argued to fulfill the Mermin-Wagner theorem by restricting the Hubbard model double occupancy to be the same as that of the impurity model which is solved numerically in an exact manner. \cite{Peters_etal_2019} 

\subsection{Causality}
DMFT, CDMFT, and DCA can be shown to be causal. We are not aware of a proof of causality for dual fermion, dual boson, dynamical vertex approximation, parquet, or FRG \cite{m_hettler_98a,m_hettler_00a}. However, in practice, that does not seem to be a serious problem.

\subsection{Crossing symmetry, Pauli principle}
\begin{figure}[htbp]
\includegraphics[height=1.80cm, width=6.0cm]{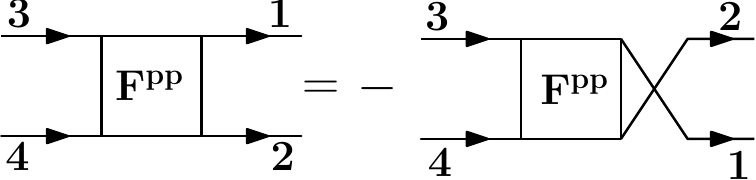}
\caption{Diagram illustration of the crossing symmetry for the particle-particle vertex.
} 
\label{fig:crossingSymmetry}
\end{figure}

An  important feature of the parquet method is that the crossing symmetries, as illustrated for the particle-particle vertex in Fig.~\ref{fig:crossingSymmetry}, expressing the identity: $$F^{pp}(12,34)=-F^{pp}(21,34),$$ are satisfied. The full FRG also satisfies the crossing symmetries. However, in practical implementations, the crossing symmetries may sometimes be broken due to the approximations made on the vertex functions.

\section{Applications}
\label{sec:Applications}

In this section we discuss the major applications of the multiscale many body methods discussed above. The list of applications here is not intended to be exhaustive, in particular we focus exclusively on studies of lattice models. Efforts to combine density functional theory with these post-DMFT/DCA methods are the subject of current active research. Readers interested in these efforts will find useful discussions in references [\onlinecite{Lichtenstein_2013,Boehnke_Lechermann_2012,Lechermann_etal_2017,Galler_etal_2019,Galler_etal_2018,Galler_etal_2017,Tomczak_etal_2017,Held_etal_2011,Schuler_etal_2017,Biermann_Lichtenstein_2017,a_toschi_11}].

\subsection{Parquet}

The parquet formalism is quite demanding from the point of view of computational resources and, for this reason, it has not been as extensively pursued as other diagrammatic methods despite its promise. We highlight two earlier studies using this method. Recent studies include applying it to the dual fermion diagrams and dynamical vertex approximation, or simplifying the calculation by different representations of the vertex functions as discussed in previous subsections. 

\begin{figure}[htbp]
\includegraphics[height=8.0cm, width=7.0cm]{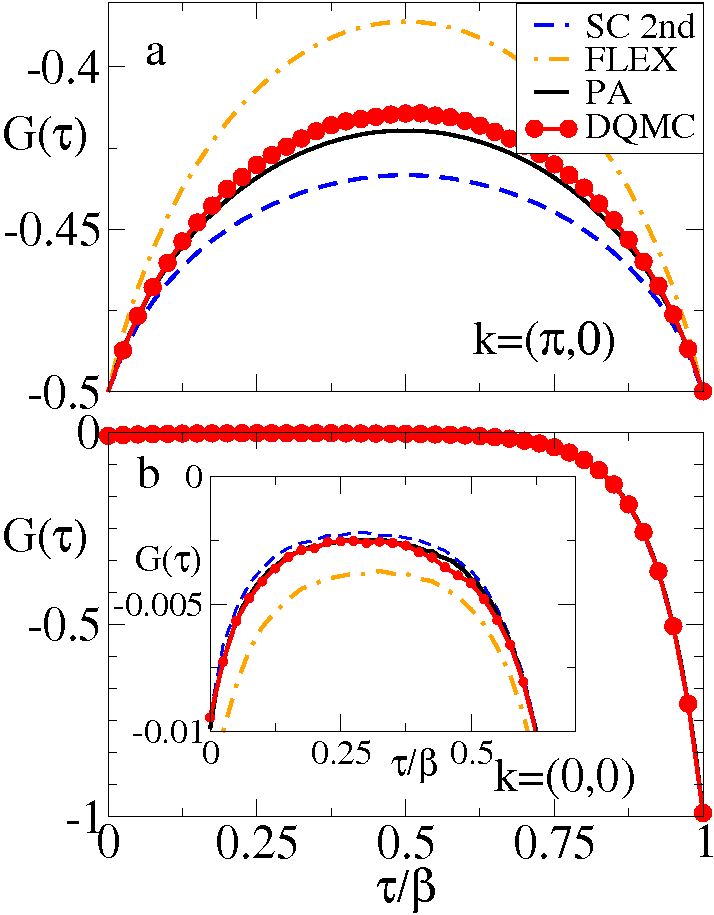}
\caption{Comparison of the Determinant Quantum Monte Carlo (DQMC) imaginary time Green function at $k=(\pi, 0)$ [panel (a)] and $k=(0,0)$ [panel (b)] with the result from several diagrammatic approaches [self-consistent second order (SC 2nd), fluctuation exchange (FLEX), parquet approximation (PA)]. The parquet approximation shows the best agreement with the DQMC result. From Ref.~[\onlinecite{Yang_etal_2009}].
} 
\label{fig:parquetComparison}
\end{figure}

S. X. Yang et al. solved the full parquet equation for the half-filled 4x4 cluster. \cite{Yang_etal_2009} Results were compared to those obtained from the Determinant Quantum Monte Carlo (DQMC), Fluctuation Exchange (FLEX), and self-consistent second-order approximation methods. This comparison, illustrated in Fig.~\ref{fig:parquetComparison}, shows a satisfactory agreement with DQMC and a significant improvement over the FLEX or the self-consistent second-order approximation.

When exploring the origin of instabilities identified in the susceptibility, the parquet formalism enables the identification of separate channel contributions. Indeed, a key feature of the parquet equations is that they express the contribution to a given scattering channel by processes from other channels. This is essential in understanding the mechanisms for key processes and phase transitions. Yang et al analyzed the pairing vertex as a function of temperature and doping. To this end, they wrote it in the form of its contributions from the charge and spin channels: $\Gamma=\Lambda+\Phi_{c}+\Phi_{s}$, and applied the $d$-wave projection to this equation to get the expression in terms of the different components: $V_{d}=V_{d\Lambda}+V_{d c}+V_{d s}$. This analysis indicated that the dominant contribution to $V_d$ originates from the spin channel.\cite{S_Yang_2011}  



\begin{figure*}[htbp]
\includegraphics[height=11.0cm, width=5.50cm]{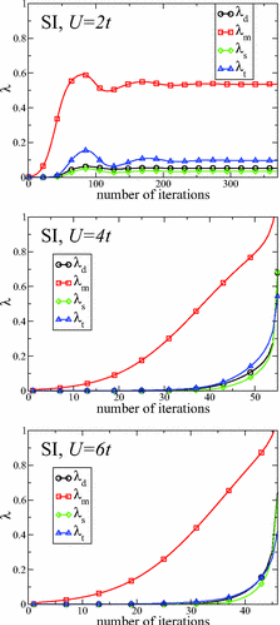}
\includegraphics[height=11.0cm, width=5.50cm]{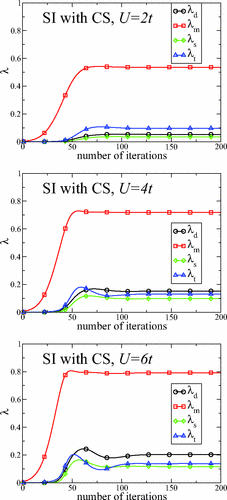}
\caption{For a simple iterative solution of the parquet formalism, leading eigenvalues of the pairing matrix in different channels [density (d), magnetic (m), spin singlet (s), and spin triplet (t)] as a function of the iteration number at temperature $T=0.4t$ on a $2\times2$ cluster. Without enforcing the crossing symmetries, the leading eigenvalues approach $1$ and divergence occurs prematurely in the self-consistent solution for $U=4t$ and $U=6t$. When the crossing symmetries on $F$ are explicitly restored at every iteration, the solution is found to remain stable for the same values of $U$. From Ref.~[\onlinecite{k_tam_13}]. 
} 
\label{fig:CSinstabilityFix}
\end{figure*}

The relations (\ref{eq:F_d}, \ref{eq:F_m}, \ref{eq:F_s}, \ref{eq:F_t}) can also allow us to restore the crossing symmetries for the full vertices during the iterative solution of the self-consistent equations. This procedure was found to enhance the stability of the self-consistent solution. With two-particle diagrams, physical instabilities can be identified by examining the divergence of the susceptibility in the associated scattering channel. This divergence is also manifested through the leading eigenvalues of the pairing matrix in the channel $r$, $M_r = \Gamma_r \times \chi_0^r$, becoming equal to $1$. As shown in Fig.~\ref{fig:CSinstabilityFix},  the leading eigenvalues of the pairing matrix typically diverge prematurely in the self-consistent solution of the parquet formalism and thus lead to a breakdown of the numerical solution at moderate to large values of the interaction. However, when the crossing symmetries are enforced throughout the iterative process, a more stable solution is found. This allows a better approach to the actual physical instability.\cite{k_tam_13}

Kusunose solved parquet equations for both the impurity Anderson model and the Hubbard model on a square lattice mainly for the particle-hole symmetric or the half-filled case. \cite{Kusunose_2010} He argued that in both models the vertex renormalization in the spin channel eliminates the magnetic instabilities of the mean-field theory to ensure satisfaction of the Mermin-Wagner theorem. The parquet method gives the same critical exponents as the self-consistent renormalization Moriya theory in the quantum critical region.

Pudleiner et al. studied the Pariser-Parr-Pople (PPP) model or Hubbard model with nonlocal interaction for the conjugated $\pi$ bonds in benzene. \cite{Pudleiner_etal_2019} They found that quasiparticle renormalization is much weaker in the PPP than in the Hubbard model, but the static part of the self-energy enhances the band gap of the PPP model. In addition, the vertex corrections to the optical conductivity are much more important in the PPP model.

\subsection{Dynamical vertex approximation}

Most of the applications of the dynamical vertex approximation are focused on the Hubbard model in two dimensions. The one dimensional Hubbard model has also been studied by the full parquet dynamical vertex approximation. \cite{Valli_etal_2015} 
For the three dimensional Hubbard model, it has been found that the antiferromagnetic phase develops incommensurate magnetic ordering as the doping increases. \cite{Schafer_etal_2017} Interesting results on the critical exponent have also been obtained. \cite{Rohringer_etal_2011} 

The main gain of including more spatial fluctuations is the suppression of the tendency towards ordering. The dynamical vertex approximation has been shown to reduce the transition temperature of the antiferromagnetic ordering 
in the half-filled Hubbard model in both two and three dimensions.
As we discussed in the previous section, neither DMFT nor dynamical vertex approximation fulfill the Mermin-Wagner theorem, thus broken symmetry is allowed at a finite temperature even for the two-dimensional case. We summarize the major  results from the dynamical vertex approximation in the following.

\subsubsection{Hubbard Model}
\begin{figure}[htbp]
\includegraphics[height=8.50cm, width=8.50cm]{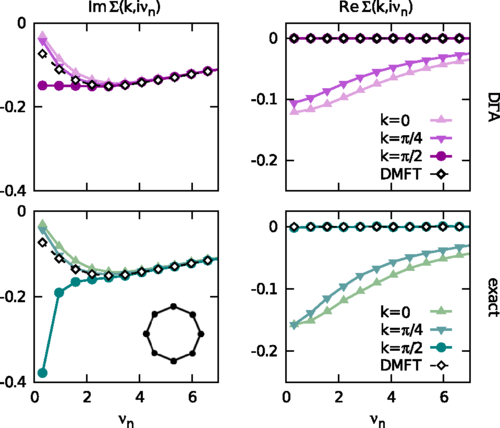}
\caption{Nonlocal corrections to DMFT obtained by the parquet D$\Gamma$A on an 8 site Hubbard ring at half-filling for $U = 2t$ and temperature $T=0.1t$. The figure shows the comparisons between the exact, and the D$\Gamma$A results for the real and the imaginary parts of the self-energy at different momentum points and for the DMFT results as a function of Matsubara frequency. From Ref.~[\onlinecite{Rohringer_et_al_RMP2018}].} \label{fig:DGammaAcomparison}
\end{figure}

Valli et al. used the parquet dynamical vertex approximation to study the electronic self-energies and the spectral properties of the finite-size one-dimensional Hubbard model with periodic boundary conditions. In this model the Fermi liquid theory is invalid, and that should present a rather challenging case for any perturbative expansion on top of the DMFT solution.\cite{Valli_etal_2015} 
Valli et al. suggested that for a non-degenerate bare dispersion, the parquet dynamical vertex approximation quantitatively reproduces the exact many-body solution of the system. This is illustrated in Fig.~\ref{fig:DGammaAcomparison} that compares the exact self-energy to that of the parquet D$\Gamma$A. Given that the system should be a Luttinger liquid which cannot be adiabatically tuned into a Fermi liquid, this is a very encouraging result.  

Sch\"{a}fer et al. have studied the two-dimensional Hubbard model on a square lattice. \cite{Schafer_etal_2016} They defined two transition lines in the phase diagram: one for the gap cutting across the nodal direction and the other for a gap throughout the Fermi surface. The self-energy data shows that the evolution between the two regimes occurs in a gradual way, not through a phase transition, and also that at low enough temperatures the whole Fermi surface is always gapped.

Sch\"{a}fer et al. also showed that the electron self-energy is well separable into a local dynamical part and a static nonlocal contributions for the three dimensional Hubbard model. \cite{Schafer_etal_2015a} The quasiparticle weight remains essentially momentum independent for different fillings, including in the presence of overall large nonlocal corrections to the self-energy.

Pudleiner at al. computed the self-energy for the half-filled Hubbard model on a square lattice using lattice quantum Monte Carlo simulations and the dynamical vertex approximation. \cite{Pudleiner_etal_2016} The self-energy is strongly momentum-dependent, but it can be parametrized via the noninteracting energy-momentum dispersion $\epsilon_k$, except for some pseudogap features right at the Fermi edge.

In Ref. \onlinecite{Schafer_etal_2015b}, Sch\"{a}fer et al. studied the two dimensional Hubbard model by combining dynamical vertex approximation, lattice quantum Monte Carlo, and variational cluster approximation. 
They demonstrated that scattering at long-range fluctuations due to paramagnons opens a spectral gap at weak-to-intermediate couplings, irrespective of the preformed localized or short-ranged magnetic moments. They argued that the two-dimensional Hubbard model has a paramagnetic phase which is insulating at low enough temperatures for any finite interaction and no Mott-Hubbard transition is observed.

Sch\"{a}fer et al. found that the antiferromagnetic phase transition of the Hubbard model in three dimensions is in contradiction with the conventional Hertz-Millis-Moriya theory. \cite{Schafer_etal_2017} They argued that the quantum critical behavior is driven by the Kohn anomalies of the Fermi surface, even when electronic correlations become strong.

Rohringer at al. studied the three dimensional half-filled Hubard model. \cite{Rohringer_etal_2011} They found the Neel temperature is lowered from that of the DMFT as expected. More interestingly, they found the critical exponents to be the same as those of the three dimensional Heisenberg antiferromagnet in contrast to mean field exponents. This demonstrates that non-mean-field behavior can indeed be obtained by these systematic nonlocal corrections. 

Rohringer and Toschi studied several spectral and thermodynamic properties of the Hubbard model in two and three dimensions. \cite{Rohringer_Toschi_2016} Specifically, by evaluating the electronic scattering rate and the quasiparticle mass renormalization in the low energy regime, they characterized the gradual deterioration by nonlocal correlations of the Fermi liquid physics as a function of the interaction strength. They found that the kinetic energy either increases or decreases  compared to that of the DMFT depending on the interaction strength being weak or strong, respectively. They argued that these results correspond to the evolution of the ground state from a nesting-driven (Slater) to a superexchange-driven (Heisenberg) antiferromagnet.

\subsubsection{Attractive Hubbard Model}
Lorenzo Del Re et al. studied the attractive Hubbard model in three dimensions for the pairing or charge density wave ordering. \cite{DelRe_2019} They found that the fitted critical exponents from the ladder D$\Gamma$A results were larger not only than the DMFT ones, but also larger than the exact ones belonging to the corresponding universality class.

\subsubsection{Periodic Anderson Model}
Sch\"{a}fer et al. studied the phase diagram and quantum critical region of the periodic Anderson model. \cite{Schafer_etal_2019} They found a phase transition between a zero-temperature antiferromagnetic insulator and a Kondo insulator. In the quantum critical region, they determined a critical exponent $\gamma=2$ for the antiferromagnetic susceptibility. This becomes $\gamma=1$ at high temperature.

\subsection{Dual Fermions} 

The dual Fermion method has been used to investigate models from the Falicov-Kimball to the Hubbard model in both two and three dimensions. The Anderson model for random disorder has also been studied and so far, this is the only post-DMFT method for the study of disorder. Perhaps a more interesting study is that of the Anderson-Hubbard model, which investigates the long standing problem of the competition between Mott insulator and Anderson insulator.  We summarize the major results from the dual fermion method in the following.

\subsubsection{Hubbard Model}
Hafermann et al. used the ladder diagrams for the dual fermion and found that the critical N\'{e}el temperature of the mean-field solution is suppressed in the ladder approximation of the two-dimensional Hubbard. \cite{Hafermann_etal_2009}

Rubtsov at al. found that the antiferromagnetic pseudogap, the Fermi-arcs formation, and the non-Fermi-liquid effects due to the Van Hove singularity are correctly reproduced by the lowest-order diagrams for the two-dimensional Hubbard model. \cite{Rubtsov_etal_2009}

Otsuki et al. obtained the phase diagram for the two-dimensional Hubbard model. \cite{Otsuki_etal_2014} This features a phase separation region in the low-doping regime around the Mott insulator. 

Astretsov et al. mapped out the phase diagram of the 2D Hubbard model as a function of temperature and doping. They identified an antiferromagnetic region at low doping and a superconducting dome at higher doping. \cite{Astretsov_etal_2020} Their results support the role of the van Hove singularity as an important ingredient for the high value of $T_c$ at optimal doping. At small doping, the destruction of antiferromagnetism is accompanied by an increase of the charge fluctuations supporting the scenario of a phase-separated state driven by quantum critical fluctuations.

Tanaka studied the square-lattice Hubbard model at half-filling using the ladder dual fermion approximation. He found that the almost simultaneous creation of the pseudogap and the loss of the Fermi liquid feature is consistent with what is expected in the Slater regime. \cite{Tanaka_2019} Although the pseudogap still appears in the quasi-particle-like single peak for $U \leq 4$, the Fermi-liquid feature is partially lost on the Fermi surface already at higher temperatures as expected in the Mott-Heisenberg regime, where local spins are preformed at high temperatures. A sharp crossover from a pseudogap phase to a Mott insulator at finite $U \simeq 4.7t$ was found to occur below the temperature of the pseudogap formation.

van Loon et al. applied the dual fermion approach with a second-order approximation to the self-energy for the Mott transition in the two-dimensional Hubbard model. \cite{vanLoon_etal_2018} A strong reduction of the critical interaction and an inversion of the slope of the transition lines with respect to single-site dynamical mean-field theory was observed. 

Katanin et al. showed the suppression of the quasiparticle weight in the three-dimensional Hubbard model. \cite{Katanin_etal_2009} With an additional correction in the susceptibility to fulfill the Mermin-Wagner theorem, \cite{Moriya_book} they also found a dramatically stronger impact of spin fluctuations in two dimensions where the pseudogap is formed at low enough temperatures. They proposed that the origin of the pseudogap at weak-to-intermediate coupling is in the splitting of the quasiparticle peak.

Hirschmeier et al. studied the three dimensional Hubbard model and they reported that in the weak-coupling regime, spin-flip excitations across the Fermi surface are important while the strong-coupling regime is described by Heisenberg physics. \cite{Hirschmeier_etal_2015} For intermediate interaction, aspects of both local and nonlocal correlations appear. They also found that the critical exponents of the transition in the strong-coupling regime are consistent with the Heisenberg model down to an interaction of $U=10t$. Again the identification of non mean-field exponents is an interesting finding. 

Antipov et al. demonstrated that diagrammatic multiscale methods anchored around local approximations are indeed capable of capturing the non-mean-field nature of the critical point of lattice models.\cite{Antipov_etal_2014} This is an interesting result as the mean field theory describes the longest length scale in the problem.

\begin{figure}[htbp]
\includegraphics[height=6.0cm, width=8.0cm]{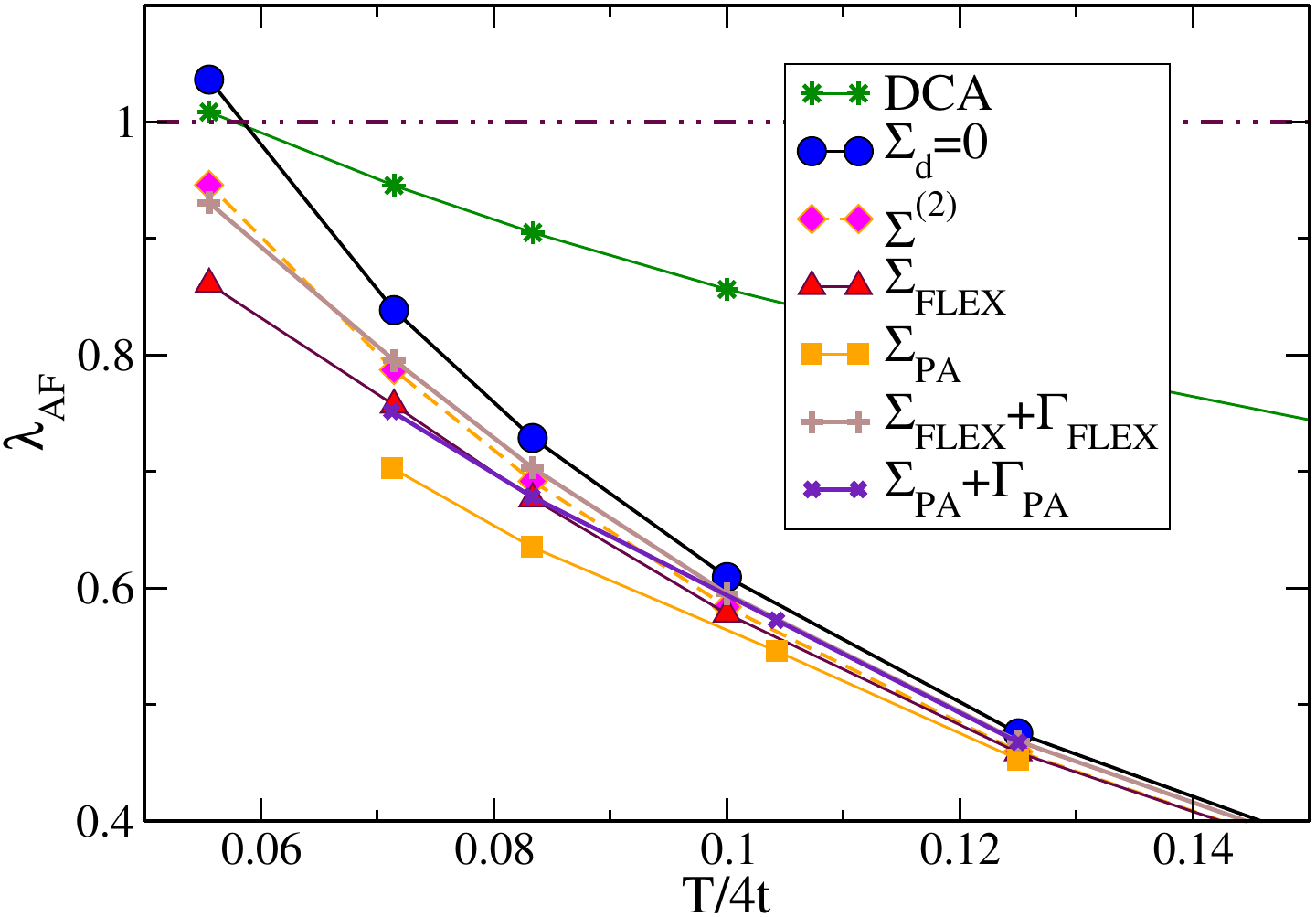}
\caption{Leading eigenvalues of the antiferromagnetic pairing matrix as a function of temperature for $U=4t$ for the DCA 1 site cluster and the dual fermion corrected result on a $4\times4$ cluster. The dual fermion corrections are calculated with different approximations: second order $\Sigma^{(2)}$, FLEX, and parquet. The leading eigenvalue is calculated by either including only the single particle correction in the self-energy, or both the single particle correction and the two-particle correction in the irreducible vertex. $\Sigma_d=0$ corresponds to the bare dual fermions quantities with no DCA calculation. From Ref.~[\onlinecite{S_Yang_2011_DFDCA}].
} 
\label{fig:eigValCorrections}
\end{figure}

van Loon et al. studied the two-dimensional square-lattice for small to moderate interaction strengths. \cite{vanLoon_etal_2018b} The nonlocal correlations beyond dynamical mean-field theory induce a pseudogap in the density of states. The upper bounds on the crossover temperature are found to be significantly lower than previously reported dynamical vertex approximation results at U=t.

As mentioned previously, the methods presented here can be applied to quantum cluster theories such as DCA and CDMFT to perturbatively capture nonlocal correlations beyond the length scale of the initial cluster size. Fig.~\ref{fig:eigValCorrections} shows an analysis of the leading eigenvalue of the antiferromagnetic pairing matrix for different approximate methods within the dual fermion approach. The figure indicates how the unphysical phase transition obtained from the mean field result (DCA with one site) is suppressed with different approximations of the dual fermion solution.\cite{S_Yang_2011_DFDCA} 

\begin{figure}[htbp]
\includegraphics[height=7.0cm, width=8.0cm]{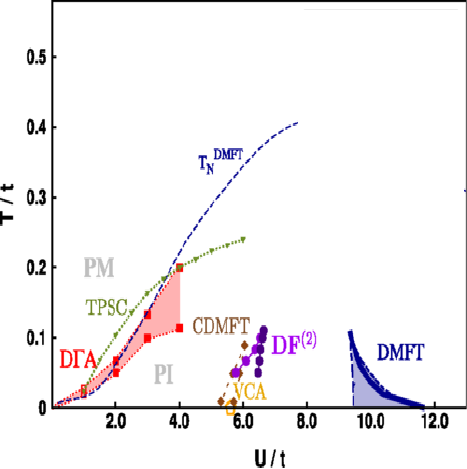}
\caption{Comparative phase diagrams of the two-dimensional Hubbard model obtained from different methods. The figure demonstrates the effects of systematic incorporation of nonlocal correlations on the Mott metal insulator transition temperature as a function of interaction. From Ref.~[\onlinecite{Rohringer_et_al_RMP2018}].} 
\label{fig:phaseDiagramMSMBcomparisons}
\end{figure}

Fig.~\ref{fig:phaseDiagramMSMBcomparisons} from Ref~[\onlinecite{Rohringer_et_al_RMP2018}] shows a compilation of results obtained using different methods and illustrates the systematic corrections to the transition temperatures through the incorporation of nonlocal correlations in the two-dimensional Hubbard model at half-filling. The figure depicts, at large $U$ and at low $T$, the DMFT paramagnetic metal solution indicating the  first-order Mott metal insulator transition with a low temperature Mott paramagnetic insulator. The first-order transition terminates at a critical value of $U_c = 10t$. Including the short range antiferromagnetic correlations such as in CDMFT, variational cluster approximation (VCA)  or second-order dual fermions (DF$^{(2)}$), modifies the critical interaction value and the shape of the coexistence region. Including longer-range antiferromagnetic fluctuations through ladder D$\Gamma$A or the two-particle self-consistent (TPSC) method leads to further modifications eventually transforming the MIT into a crossover at small $U$ that is consistent with the $U_c \; \to \; 0$ for $T \; \to \; 0$ limit.

\subsubsection{Hubbard Model on a Triangular Lattice}

Yudin at al. studied the Hubbard model on a triangular lattice. \cite{Yudin_etal_2014} They showed that the band flattening is driven by correlations and is well pronounced even at sufficiently high temperatures, of the order of 0.1–0.2 times the hopping parameter. 

Lee et al. studied the Hubbard model on the triangular lattice at half filling. They determined the metal-insulator transition and the hysteresis  associated with a first-order transition in the double-occupancy and nearest-neighbor spin-correlation functions as functions of temperature. \cite{Lee08}  By calculating the spin susceptibility, an enhancement of antiferromagnetic correlations and  evidence for magnetically ordered phases were found.

Antipov et al. studied the half-filled Hubbard model on an isotropic triangular lattice with a spin polarized extension of the dual fermion approach. \cite{Antipov_etal_2011} They found that the dual fermion corrections drastically decrease the energy of a spin liquid state while leaving the non-collinear magnetic states almost non-affected. This makes the spin liquid become a preferable state in a certain interval of interaction strength of the same order of magnitude than the bandwidth.

Li et al, studied both the half-filled and the doped Hubbard model on a triangular lattice and produced its phase diagram. \cite{Li_etal_2015} 

\subsubsection{Hubbard Model on the Honeycomb Lattice}
Hirschmeier et al. studied the Hubbard model on the honeycomb lattice in the vicinity of the quantum critical point by means of a multiband formulation of the dual fermion approach. They found that the critical interaction strength of the quantum phase transition from a paramagnetic semimetal to an antiferromagnetic insulator is in good agreement with other numerical methods. \cite{Hirschmeier_etal_2018} They also argued that the Hubbard model on the honeycomb lattice behaves like a quantum nonlinear $\sigma$ model, while displaying signs of non-Fermi-liquid behavior.

\subsubsection{Falicov-Kimball Model}
Astleithner at al. studied the Falicov-Kimball model.  Using the full parquet dynamical vertex approximation, they argued that weak localization corrections in the particle-particle channel are not the dominant vertex corrections to the optical conductivity. \cite{Astleithner_etal_2019}

\subsubsection{Kondo Lattice Model}

Otsuki studied the Kondo lattice model to explore possible superconductivity emerging from the critical antiferromagnetic fluctuations. \cite{Otsuki_2015} The d-wave pairing is found to be the leading instability only in the weak-coupling regime. As the coupling is increased, a change of the pairing symmetry into a p-wave spin-singlet pairing was found.

\subsubsection{s-d Exchange Model}
Sweep et al. studied the critical values of the s-d exchange coupling constant. \cite{Sweep_etal_2013} They reported a difference between the DMFT and dual fermion results that is more than a factor of two for the square lattice and spin one-half localized electrons.

\begin{figure}[htbp]
\includegraphics[height=9.0cm, width=8.0cm]{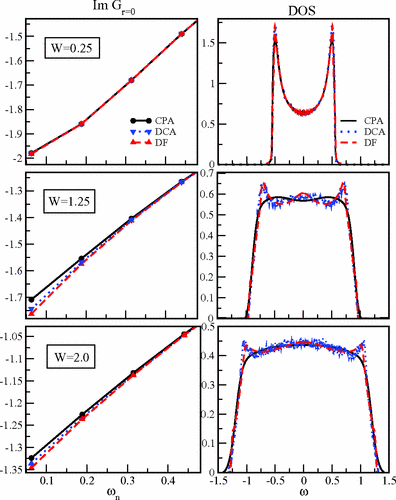}
\caption{Nonlocal corrections to the CPA for the disorder Anderson model. The dual fermion corrections capture the nonlocal corrections that are absent in the CPA results. This leads to a better agreement with the DCA result for the imaginary part of the self-energy and the density of states as the disorder strength is increased. Left panel: $Im\mathrm{Gr_{=0}}(\omega_n)$ in $d=1$ at $T=0.02$. Right panel: total density of states for different disorder strengths: $W=0.25, \; 1.25, \;2.0 \; (4t=1).$ From Ref.~[\onlinecite{h_terletska_13}].
} 
\label{fig:DualFermionCPAcorrections}
\end{figure}

\subsubsection{Anderson Disorder Model}
Terletska et al. generalized the dual fermion approach to disordered systems using the replica method. \cite{h_terletska_13,s_yang_13a} The developed method utilizes the exact mapping to the dual fermion variables, and includes inter-site scattering via diagrammatic perturbation theory in the dual variables. As shown in Fig.~\ref{fig:DualFermionCPAcorrections} nonlocal effects that are missed in the CPA are captured.

\subsubsection{Anderson Falicov-Kimball Model}
Yang at al. generalized the dual-fermion formalism for disordered fermionic systems to include the effect of interactions. The phase diagram for the two dimensional  Anderson-Falicov-Kimball model was obtained.
\cite{s_yang_13b}

\subsubsection{Anderson-Hubbard Model}
\begin{figure}
    \centering
    \includegraphics[height=16.0cm, width=7.0cm]{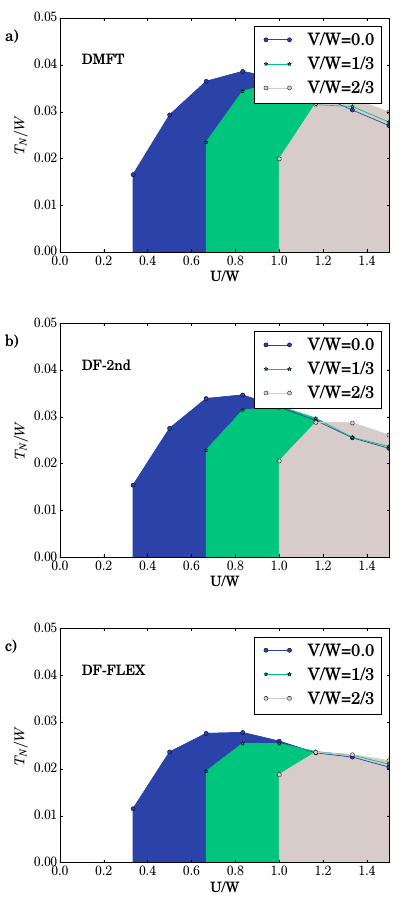}
    \caption{$U$-$T$ phase diagram of the 3D Anderson-Hubbard model for different values of the disorder strength $V$ obtained with DMFT (a), dual fermions with second order diagrams (b), and dual fermions with FLEX diagrams (c). The shaded regions correspond to the antiferromagnetic phase. $W$ is the bandwidth. Note that the authors do not calculate the critical temperatures in the region of $U/W < 1/3$. From Ref.~[\onlinecite{Haase_etal_2017}].} 
    \label{fig:U_T_phaseDiagram_3DAndersonHubbard}
\end{figure}

Haase et al. studied the three-dimensional Anderson Hubbard model. They report that the dual-fermion approach leads to quantitative as well as qualitative improvement of the dynamical mean-field results. This is shown in the phase diagrams of Fig.~\ref{fig:U_T_phaseDiagram_3DAndersonHubbard} obtained with DMFT (a), dual fermions with second order diagrams (b), and dual fermions with FLEX diagrams (c). The systematic improvement of the solution first with the incorporation of non-local corrections and then in terms of the level of the diagrammatic treatment is shown through the expected suppression of the DMFT critical temperatures for the antiferromagnetic phase. These solutions allowed the authors to calculate the hysteresis in the double occupancy in three dimensions, taking into account nonlocal correlations.\cite{Haase_etal_2017}

Otsuki studied the Kondo lattice model,\cite{Otsuki_2015} and found that different superconductivity pairing symmetries emerge from the critical antiferromagnetic fluctuations. He found the d-wave pairing to be the leading instability only in the weak-coupling regime. As the coupling is increased, a change of the pairing symmetry into a p-wave spin-singlet pairing is observed. The competing superconductivities are ascribed to a  crossover between small and large Fermi surfaces, which occurs with the formation of heavy quasiparticles.

\subsection{Dual Bosons}

The dual boson method has been applied on models with nonlocal interactions. These include the truncated long range Coulomb coupling, the nearest neighbor interaction in the extended Hubbard model, and the anisotropic dipolar coupling for cold atoms. A key difference is that those nonlocal density-density type couplings can lead to the competition between charge fluctuations and spin fluctuations which results in charge density wave ordering and possibly bond wave ordering. \cite{Nakamura_2000}

\subsubsection{Extended Hubbard Model}
Vandelli et al. proposed to use quantum Monte Carlo to sample diagrams from the dual boson theory for the extended Hubbard model. \cite{Vandelli_etal_2020} They proposed that the single-particle Green function allows one to estimate the transition point to the charge density wave phase.

\subsubsection{Hubbard Model with Dipolar Coupling}
van Loon et al. studied the Hubbard model with long-range dipole-dipole  interactions. \cite{vanLoon_etal_2015} This is an interesting model in the context of experiments with cold atoms on optical lattice. \cite{Lu_etal_2012} Besides the stripe phase and the checkerboard phase, based on their dual boson calculation they suggest that there is a novel phase with "ultralong-range" density correlations at distances of tens of lattice sites.

\subsubsection{Hubbard Model with Coulomb Coupling}
Hafermann et al. and van Loon et al. studied the polarization for the two dimensional Hubbard model with long range Coulomb coupling. \cite{vanLoon_etal_2014,Hafermann_etal_2014} They found that plasmon spectra are qualitatively different from those of the random-phase approximation: they exhibit a spectral density transfer and a renormalized dispersion with enhanced deviation from the canonical behavior.

\subsection{TRILEX}

Applications of TRILEX are mostly on the two dimensional Hubbard model. 


Aryal et al. found that the local vertex, for strong interactions, gains a strong frequency dependence, driving the system to a Mott transition for the half-filled Hubbard model on a square lattice. \cite{Ayral_Parcollet_2015} At low enough temperatures, large spin fluctuations lead to an enhancement of the momentum dependence of the self-energy. Upon doping, they find a Fermi arc in the spectral function.

Vučičević et al. studied the dependence of the superconducting temperature on the bare dispersion at weak coupling, which shows a clear link between strong antiferromagnetic correlations and the onset of superconductivity. \cite{Vucelse_etal_2017} They identified a combination of hopping amplitudes particularly favorable to superconductivity at intermediate doping.

\subsection{FRG}
Functional renormalization group has been used extensively for over two decades. Recent applications to improve the DMFT solution have so far been mostly limited to the Hubbard model.


Tranto et al. first proposed to use FRG to expand the DMFT solution. \cite{Tranto_etal_2014} They studied the half-filled square lattice Hubbard model and found that the method provides more prominent momentum dependence than the conventional FRG method. 

Vilardi et al. studied the doped two dimensional Hubbard model. \cite{Vilardi_etal_2019} They found strong antiferromagnetic correlations from half-filling to 18\% hole doping at low temperature, and a sizable d-wave pairing interaction driven by magnetic correlations at the edge of the antiferromagnetic region.

\section{Conclusion}
We have reviewed multiscale many body numerical methods to address strongly correlated systems by appropriately treating the short length scale, the long length scale and the intermediate length scale. The different methods implemented to date have produced promising results despite being hindered by a variety of numerical challenges. Since short length scales are treated exactly, diagrammatic methods arise as a suitable approach to deal with the intermediate length scales by systematically evaluating appropriate subsets of possible diagrams. In this context the parquet formalism is the most natural toolkit. We have reviewed the construction of the parquet formalism and the different diagrammatic approximations that it encompasses as well as algorithms for their numerical solutions. We have not discussed in this review efforts to extend the methods into ab-initio calculations. These represent an important next step for appropriate treatments of real materials. In general, multiscale many body methods to incorporate nonlocal corrections into the DMFT solution represent an active area of research and new implementations are actively being developed to overcome previous shortcomings.

Some of the latest ideas have not been discussed in the present review. These include but are not limited to the parquet method for the vertex in the boson-fermion representation, \cite{Krien_etal_2020a,Krien_etal_2019,Krien_etal_2020b,Krien_etal_2020c} the atomic approximation of the four-particle irreducible functional method, \cite{Ayral_Parcollet_2016} one-particle irreducible functional method, \cite{Rohringer_etal_2013} nonlocal expansion method, \cite{Li_2015} and FLEX+DMFT approach. \cite{Kitatani_etal_2015,Kitatani_etal_2017}

Another important topic is that of the solvers for the vertex functions. While there are many different numerical solvers for the impurity/cluster problem, most of them are not suitable for the calculation of the vertex function which is essential  for perturbative expansions around the DMFT solution. While many solvers may be generalized for the calculation of vertex function, at present the practical methods are exact diagonalziaton and quantum Monte Carlo.

For the exact diagonalization method, the calculation of the vertex function is usually down to brute force calculation in the K\"{a}ll\'{e}n-Lehmann spectral representation. Unlike the calculation of the single particle quantity, the method based on expansion in terms of a continued fraction is not applicable for the calculation of the vertex functions. \cite{Lin_etal_1993,Senechal_2010,a_georges_96a,liebsch2011temperature} Thus the calculation is limited to a rather small number of bath sites or orbitals. \cite{Tanaka_2019,a_toschi_07}

For the quantum Monte Carlo approach, besides the minus sign problem, \cite{Shinaoka_etal_2015,Kim_etal_2020} the main challenge for calculating the vertex is the noise in the measurements, especially at high frequency. This is particularly acute for the hybridization expansion approach. Significant progress has been made to reduce the noise by measuring in a basis of orthogonal polynomials. \cite{Hefermann_etal_2012,Hafermann_2014,Gull_etal_2018,Huang_2016}
There is continuous improvement on the sampling efficiency and on the ability to attain ergodicity. \cite{Semon_etal_2014a,Semon_etal_2014b,Gunacker_etal_2015,Melnick_etal_2020,Gunacker_etal_2016}
For further discussion of these approaches, we refer the interested readers to the comprehensive review  by Gull et al. \cite{e_gull_11}

As we have seen, the different implementations of multiscale many body approaches have produced very significant results that validate the motivation of the approach. Indeed, more appropriate treatments of nonlocal correlations improve the results both qualitatively and quantitatively. To improve the robustness of the approach and to extend the methods to broader ranges of parameters, further developments are needed to overcome the computational challenges. This may involve new insights on the physics, leading to modified algorithms, or the development of new numerical techniques.

While we discuss several executions of the MSMB approach, an omission in this paper is a definite guideline with pros and cons of the respective methods. In particular, it is desirable to answer the question of which method provides the best results with the least numerical effort. Presently, there are various reasons why it is rather difficult to address this question. First, many of the methods have not been fully investigated, some of them may not even have been optimally implemented. Second, the question of 'best results' needs qualification, it is unlikely that there is one method which holds a clear advantage over the others in terms of getting the best results. This can be understood from the point of view that all the methods discussed are based on some form of perturbative expansion on top of effective interacting models. The range of parameters is an important factor in deciding the quality of different expansions. Third, the implementation of a given method also affects the quality of the results. All the methods require the handling of different types of vertex functions. The procedure for storing and approximating the vertex functions can be a non negligible factor in the final results. While there is intense activity on the MSMB approaches, the field is still rather young. We have painted a detailed picture of the landscape in our discussions of the different methods, the nuances within the methods, and possible subtleties across the numerical approximations involved. In time, we believe the community will push these different implementations to the point of producing a fuller picture; allowing for more transparent comparisons.\\ 

\section*{Acknowledgments} 
We are deeply indebted to Mark Jarrell for introducing us to the computational studies of strongly correlated systems. We thank our collaborators over the years, Shuxiang Yang, Hanna Terletska, N.S. Vidhyadhiraja, Yi Zhang, Cyrill Slezak, Zi-Yang Meng, Harmut Hafermann, Muhammad Aziz Majid, Karlis Mikelsons, Ehsan Khatami, Peng Zhang, Peter Reis, Kuang-Shing Chen, Chinedu Ekuma, Ryky Nelson, Patrick Haase, Nagamalleswararao Dasari, Sheng Feng, Samuel Kellar, Brian Moritz, Paul Kent, Unjong Yu, Dimitris Galanakis, Hartmut Kaiser, Karen Tomko, Thomas Maier and Thomas Pruschke. We thank Eric Dohner for proofreading our manuscript.

HF is partially supported by NSF PHY-2014023.   
KMT and JM are partially supported by  the U.S. Department of
Energy, Office of Science, Office of Basic Energy Sciences under Award Number DE-SC0017861.
KMT is also supported by NSF DMR-1728457 
and OAC-1931445.  

\bibliographystyle{apsrev4-1}
\bibliography{master}




\end{document}